\documentclass[fleqn,usenatbib]{mnras}

\usepackage{newtxtext,newtxmath}

\usepackage[T1]{fontenc}

\DeclareRobustCommand{\VAN}[3]{#2}
\let\VANthebibliography\thebibliography
\def\thebibliography{\DeclareRobustCommand{\VAN}[3]{##3}\VANthebibliography}

\usepackage{graphicx}	
\usepackage{amsmath}	
\usepackage{xcolor}
\usepackage[dvipsnames]{xcolor}
\usepackage{graphicx}
\usepackage{natbib}
\usepackage{comment}
\usepackage{subfig}
\usepackage{xspace}
\usepackage{soul}
\usepackage[normalem]{ulem} 

\newcommand{\carlo}[1]   {\textbf{{\color{red}{Carlo: #1}}}}
\newcommand{\rel}{$R_d{-}r_0$\xspace}
\newcommand{\kpc}{\mathrm{kpc}\xspace}

\title[Feedback driven interactions]{Feedback driven interactions between dark and luminous matter to explain tight galaxy scaling relations. }

\author[Alrawas et al.]{
Leen Alrawas,$^{1,2,3}$\thanks{E-mail: laa9597@nyu.edu}
Andrea V. Macci\`o,$^{1,2}$
Carlo Cannarozzo,$^{1,2}$
\\
$^{1}$New York University Abu Dhabi, PO Box 129188, Saadiyat Island, Abu Dhabi, UAE\\
$^{2}$Center for Astrophysics and Space Science (CASS), New York University Abu Dhabi, Saadiyat Island, PO Box 129188, Abu Dhabi, UAE\\
$^{3}$Department of Physics, New York University, New York NY 10003, USA
}

\date{Accepted 2026 May 12. Received 2026 April 18; in original form 2025 December 8}

\pubyear{\the\year{}}

\begin{document}
\label{firstpage}
\pagerange{\pageref{firstpage}--\pageref{lastpage}}
\maketitle

\begin{abstract}
The tight empirical correlation linking the stellar disk scale length $R_d$ to the dark matter scale radius $r_0$ has been proposed as potential evidence for a fundamental coupling between baryons and dark matter beyond gravity. We re-examine the physical origin of this relation using a sample of 31 galaxies drawn from the NIHAO cosmological hydrodynamical simulations, which include no dark matter–baryon interactions other than gravity and baryonic feedback processes. NIHAO naturally reproduces both the normalization and the small scatter of the observed $R_d{-}r_0$ relation at $z=0$, while showing a slightly shallower distribution. By tracking galaxies from $z=2$ to $z=0$, we identify three evolutionary classes: systems undergoing disk expansion, contraction, and quasi-static galaxies. Using a Bayesian hierarchical framework, we provide the first evolutionary characterization of the $R_d{-}r_0$ relation, tracing how its normalization, slope, and intrinsic scatter evolve across cosmic time, from $z=2$ to the present-day Universe. Together with a mild decrease in normalization (by $\sim0.07$ dex) and a flattening of the slope from $\alpha\simeq1.05$ to $\alpha\simeq0.95$, we find that the intrinsic scatter weakly decreases toward lower redshift, indicating that galaxies tend to evolve along the relation, jointly re-balancing their stellar and dark matter scales. Comparing hydro simulations with their dark matter only counterparts, we can isolate the effect of baryons and baryonic feedback on dark matter evolution. Our results indicate that stellar feedback alone can reshape the central potential and naturally establish the observed coupling between luminous and dark matter, without the need to invoke modifications to the dark sector.


\end{abstract}




\section{Introduction} 

The $\Lambda$ Cold Dark Matter ($\Lambda$CDM) paradigm has emerged as the standard cosmological model, providing a remarkably successful framework for understanding the large-scale structure and evolution of the universe. One of its most compelling achievements lies in its capacity to explain the hierarchical formation of structure, where dark matter halos serve as the gravitational scaffolding for baryonic processes that lead to galaxy formation. Although $\Lambda$CDM has proven highly effective on cosmological scales, its success in reproducing the diverse yet tightly correlated properties of galaxies on smaller, galactic scales is a subject of ongoing scrutiny \citep{del2017small, arora2023manga}.

Galaxy scaling relations, such as the Tully-Fisher relation \citep[e.g.,][]{tully1977new, mcgaugh2000baryonic}, the stellar-to-halo mass relation \citep[e.g.,][]{brook2014stellar, girelli2020stellar}, the velocity-to-brightness relation \citep[e.g.,][]{pizzella2005relation, courteau2007bulge, lelli2013scaling}, the radial acceleration relation \citep[e.g.,][]{lelli2017one, stone2019intrinsic}, and the stellar or gas mass–metallicity relation \citep[e.g.,][]{kewley2008metallicity, ma2016origin}, represent a subset of key observational constraints on models of galaxy formation and evolution. These relations imply a degree of regularity in galaxy formation that is not expected given the stochastic nature of baryonic processes such as gas accretion, star formation, and feedback from supernovae and active galactic nuclei \citep[e.g.,][]{cole1994recipe,keller2019chaos}. The existence of these correlations has often been perceived as a challenge to $\Lambda$CDM-based models \citep[e.g.,][]{navarro2000dark, dutton2016nihao, arora2023manga}, which must simultaneously account for complex, non-linear baryonic physics and the underlying dynamics of dark matter.

In recent years, cosmological hydrodynamical simulations have made impressive progress in reproducing key galaxy scaling relations like the Tully-Fisher relation and the size–mass relation \citep{Crain2015,dutton2017nihao}, the mass metallicity relations \citep{DeRossi2017, Nelson2019, buck2021}, the stellar mass-halo mass relation \citep{wang2015nihao, pillepich2018first,  maccio2020} and many others \citep[see][for more details]{Vogelsberger2020, arora2023manga}.


Among the many observed galaxy scaling relations, there is one that is of particular interest for dark matter theories and cosmology. This is the observed correlation between the characteristic size of the dark matter halo, typically quantified by the dark matter scale radius $r_0$, and the stellar disk scale length $R_d$ \citep{donato2004cores}. This relation provides a direct link between the visible and invisible sides of galaxy formation and might shed some light on the properties of dark matter. Observational studies have revealed a remarkably tight linear relationship between these two quantities across a diverse range of galaxy morphologies, encompassing low surface brightness galaxies \citep{di2019universal}, dwarf disk galaxies \citep{karukes2017universal}, normal spiral galaxies \citep{donato2004cores}, and even extending to giant central dominant galaxies, such as M87 \citep{de2022accurate}. 

This pronounced correlation, together with the small scatter, has been interpreted by some authors \citep[e.g.][]{salucci2020paradigms} as possible evidence for a coupling between the baryonic and dark matter components that may transcend the expectations of purely gravitational interaction \citep{foot2014tully, nesti2023quest}.

On the other hand, recent simulations have consistently shown that baryonic processes (e.g., gas accretion and feedback) can alter the DM distribution in galaxies by either contracting or expanding the dark matter halo \citep{dutton2007revised, abadi2010, maccio2012ApJ, tollet2016nihao, chan2015}. This raises the question whether such feedback-driven 'interactions' can be the core mechanism to set the $R_d{-}r_0$ relation and drive its evolution through cosmic time, without the need to invoke new physics in the dark sector. 

In this work, we employ the NIHAO (Numerical Investigation of a Hundred Astrophysical Objects) simulations \citep{wang2015nihao} suite to test to what extent numerical simulations are capable of reproducing the observed relations and their scatter, in a model where no additional interactions besides gravity are included.

This paper is organized as follows: Section~\ref{simulations} provides a brief summary of the NIHAO simulations. In Section~\ref{relation}, we present the \rel scaling relation derived from observational data alongside the corresponding relation obtained from simulated galaxies. Section~\ref{tracks} discusses two distinct evolutionary pathways through which galaxies may converge toward this relation by redshift $z=0$. The temporal evolution of the scaling relation from $z=2$ to $z=0$ is examined in Section~\ref{time_evolution}. Finally, the principal findings of this work are summarized in Section~\ref{conc} together with possible caveats in Section~\ref{caveats}. 

\section{The NIHAO simulations} \label{simulations}
The NIHAO simulations \citep{wang2015nihao, blank2019nihao} used in this study are based on the \texttt{GASOLINE2} hydrodynamical code \citep{wadsley2017gasoline2}. Simulations include Compton cooling, photoionization, and heating from the ultraviolet background, metal-line cooling, chemical enrichment, star formation, and stellar feedback—including supernovae and early feedback from massive stars \citep{stinson2013making}.

The cosmological parameters of the simulation suite are consistent with the Planck 2016 data release, namely the Hubble parameter is $H_0 = 67.1\,\mathrm{km\,s}^{-1}\,\mathrm{Mpc}^{-1} \), with matter density $\Omega_m = 0.3175$, dark energy density $\Omega_\Lambda = 0.6824$, baryon density $\Omega_b = 0.0490$, power spectrum normalization $\sigma_8 = 0.8344$ and spectral index $n_s = 0.9624$ \citep{ade2016planck}.
Each galaxy is resolved with at least 500,000 particles, including DM, gas, and stars. Mass and spatial resolution vary across the sample, ranging from a DM particle mass of $m_{\mathrm{dm}} = 3.4 \times 10^3\,M_\odot$ with gravitational softening $\epsilon = 100\,\mathrm{pc}$ for dwarf galaxies, up to $m_{\mathrm{dm}} = 1.4 \times 10^7\,M_\odot$ and $\epsilon = 1.8\,\mathrm{kpc}$ for the most massive systems \citep[see][]{wang2015nihao, blank2019nihao}.

The NIHAO simulations are able to successfully reproduce a wide range of galaxy properties, including the stellar-to-halo mass relation \citep{wang2015nihao}, the relation between disc gas mass and disc size \citep{maccio2016nihao}, the Tully--Fisher relation \citep{dutton2017nihao}, the diversity of galaxy rotation curves \citep{santos2018nihao, frosst2022diversity}, and the satellite mass function of the Milky Way and M31 \citep{buck2019nihao}.

To study the relations between the dark matter scale radius $r_0$ and the stellar disk radius $R_d$, we select 33 galaxies from the NIHAO with stellar and dark matter masses similar to the observed galaxies. Namely, our objects have DM halo with masses in the range $10^{11}{-}10^{12} M_\odot$, and with stellar masses larger than $2 \times 10^9\, M_\odot$. Two of the selected systems are found to go through mergers at $z=0$; for this reason, we exclude these galaxies from the analysis. Thus, our final galaxy sample consists of 31 simulated galaxies.

\section{Reproducing the $\text{R}_{\text{\lowercase{d}}}{-}\text{\lowercase{r}}_0$ relation}
\label{relation}

In observations, the disc scale parameter ($R_{\rm d}$) is normally obtained \citep[e.g.][]{salucci2000dark,karukes2017universal} by fitting the stellar surface density profile with an exponential functional form known as the Freeman disk \citep{freeman1970disks}:
\begin{equation}
    \mu(R) = \frac{M_D}{2 \pi R_d^2} e^{-\frac{R}{R_d}},
    \label{exp}
\end{equation}
where $M_D$ is the stellar disk mass and $R_d$ is the stellar disk radius of edge-on spirals. 

To recover the dark matter scale radius $r_0$, the DM density profiles are assumed to follow a cored profile (which provides a better fit for most disc galaxies \citep{salucci2020paradigms}), which is usually parametrized with either the Burkert profile \citep{burkert1995structure}
\begin{equation}
    \rho_B(r) = \frac{\rho_0}{\left(1 + \frac{r^2}{r_0^2}\right)\left(1 + \frac{r}{r_0}\right)},
    \label{burkert}
\end{equation}
or the quasi-isothermal (ISO) profile \citep{begeman1991extended}

\begin{equation}
    \rho_{\rm {ISO}}(r) = \frac{\rho_0}{\left(1 + \frac{r^2}{r_0^2}\right)}.
    \label{iso}
\end{equation}
In both profiles $\rho_0$ is the dark matter halo central density and $r_0$ the dark matter halo scale radius.

These two scale radii, describing the distribution of the visible and invisible matter, are connected to very different formation paths, since the dark matter distribution is mostly set at the time of the halo collapse \citep{Bullock01, maccio2007concentration} while the size of the stellar disc is the final result of several competing phenomena that include, among others, gas cooling, angular momentum conservation, star formation and feedback \citep[e.g.][]{MoMaoWhite1998}.

\begin{figure}
    \includegraphics[width=\columnwidth]{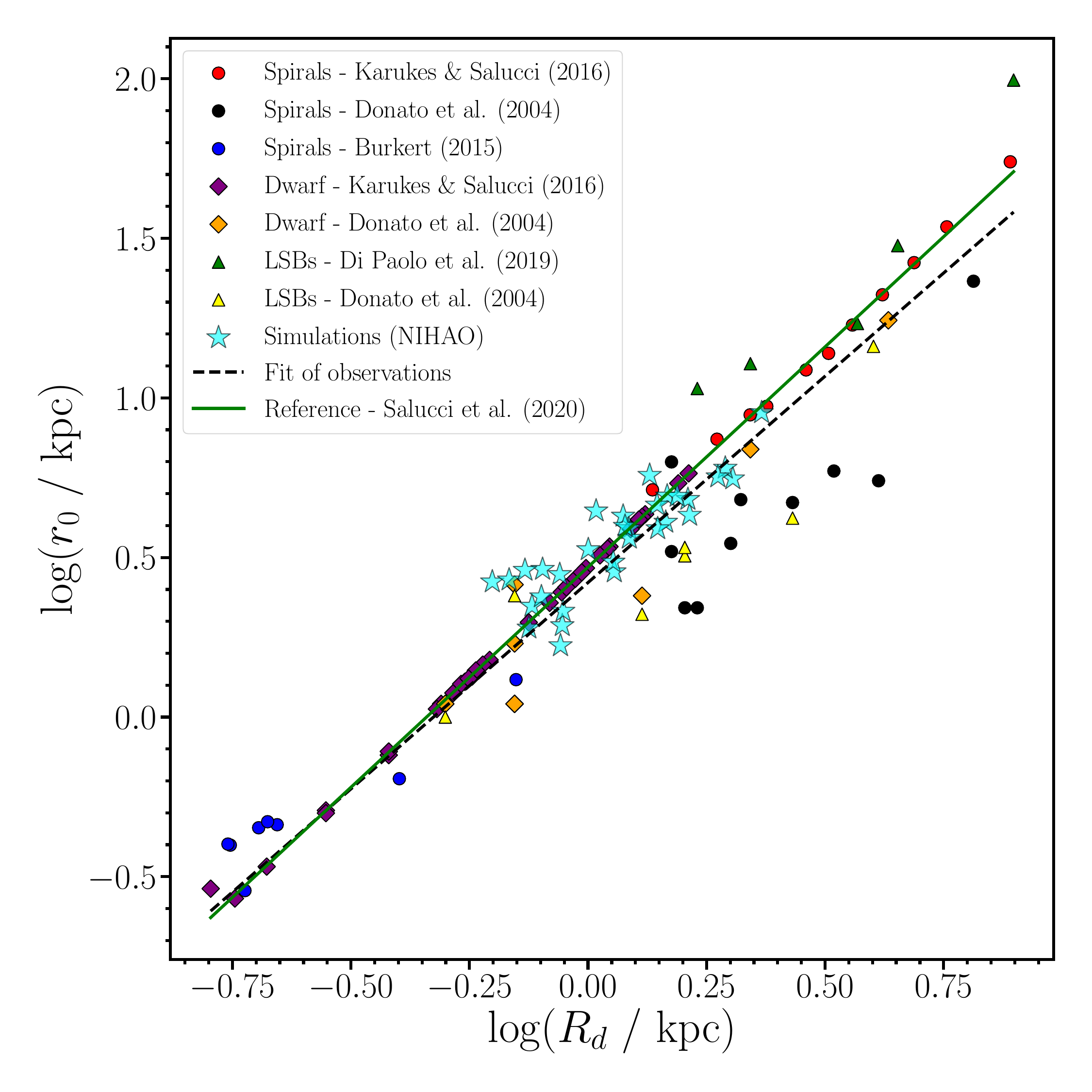}
    \caption{$r_0$ versus $R_d$ in observed and simulated galaxies. The dashed line is the fit calculated in this work for the observed galaxies, and the green line is the fit reported in \citet{salucci2020paradigms, nesti2023quest} for selected datasets.}
    \label{obs}
\end{figure}

Nevertheless, previous authors have reported a tight correlation between $r_0$ and $R_d$ with a scatter of around 0.05 (Salucci private communication). In Fig. \ref{obs} we compile the most recent results in the literature for this correlation.



\label{simulation}
Similarly to observations, we computed both the $r_0$ and $R_d$ for our set of 31 NIHAO galaxies. The dark matter scale radius $r_0$ is computed using the isothermal profile (ISO), since this profile provides a better fit to our galaxies than the Burkert profile (see Appendix \ref{appendix} for more details).

\begin{figure}
    \includegraphics[width=\columnwidth]{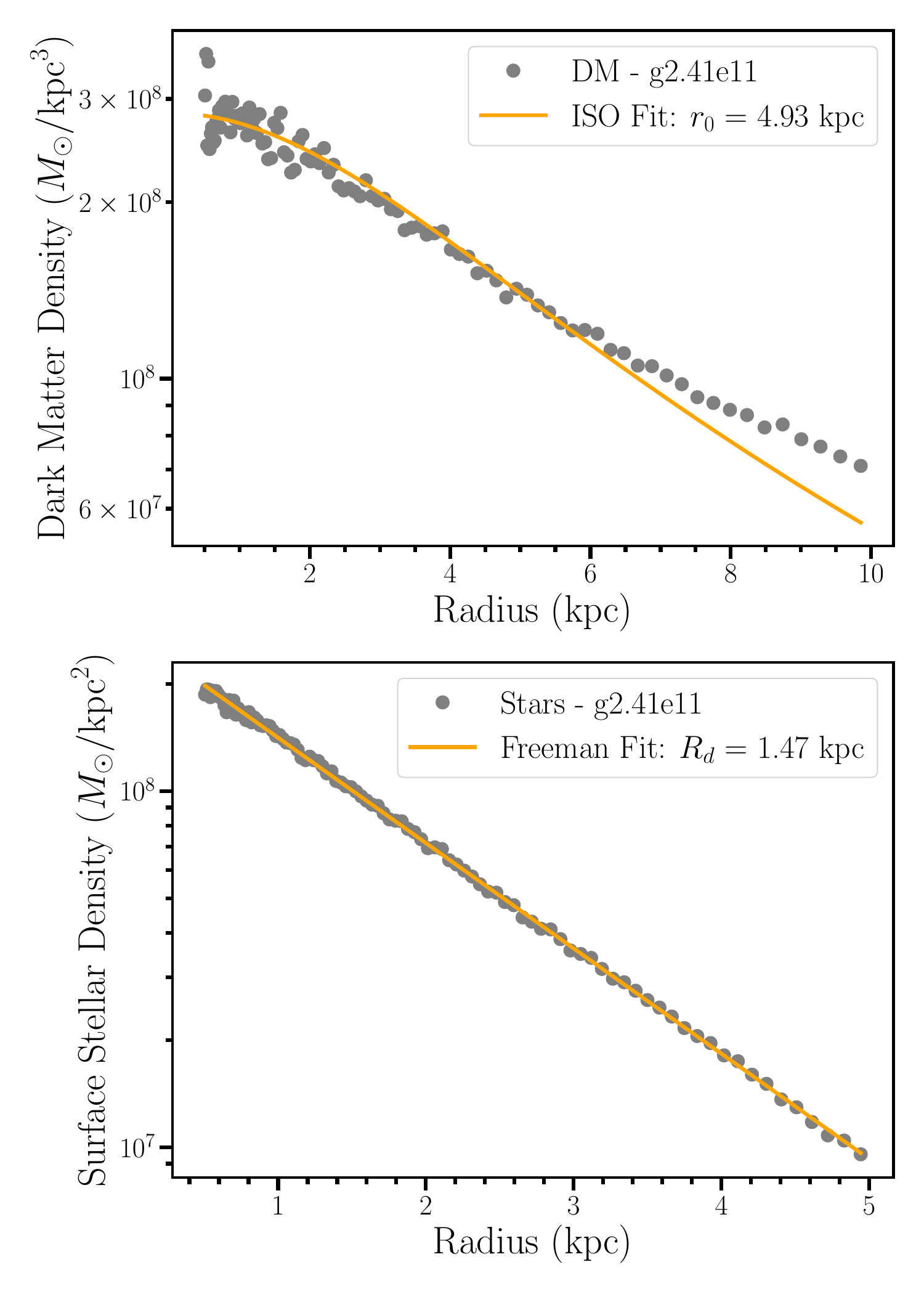}
    \caption{The dark matter density profile for a halo of mass $2.41 \times 10^{11} M_\odot$ fitted to the ISO profile \ref{iso}, and the surface stellar density profile for the same halo fitted to the Freeman disk \ref{exp}.}
    \label{fits_combined}
\end{figure}

The disk scale radius $R_{\rm d}$ is computed by fitting the projected stellar density, after rotating the galaxy edge-on\footnote{Normally disc scale lengths are better computed for face-on galaxies, but in this work we tried to follow the same procedure as in \cite{nesti2023quest} and hence we oriented our galaxies edge-on. There is an offset of about 30\% between $R_d$ values computed with the two different projections, but this does no affect any of the trends  discussed in this paper}, with the Freeman profile, following the methodology adopted in \citet{salucci2020paradigms} and \citet{nesti2023quest}. Two examples of simulated profiles and relative fits are shown in Fig.\ref{fits_combined} for the dark matter (upper panel) and the stellar surface density (lower panel) for a galaxy with total mass $2.41 \times 10^{11} M_\odot$. 

The simulation results in the $R_d{-}r_0$ plane are shown in Fig. \ref{obs} as (blue) starred symbols. Overall, our simulations are in good agreement with the observations, especially when compared with the fit of all observational data provided in this work (the dashed line).
Moreover, NIHAO galaxies also exhibit a small scatter around the relation (the total rms is 0.081 dex), in substantial agreement with observations.  

This result is noteworthy, as it demonstrates that a linear relation between the structural parameters of the dark matter halo and the stellar disc can naturally emerge in a $\Lambda$CDM Universe, without requiring additional dark matter–baryon interactions beyond gravity.


\section{Galaxy evolutionary tracks in $\text{R}_{\text{\lowercase{d}}}$-$\text{\lowercase{r}}_0$ plane} 
\label{tracks}

\begin{figure*}
    \includegraphics[width=2\columnwidth]{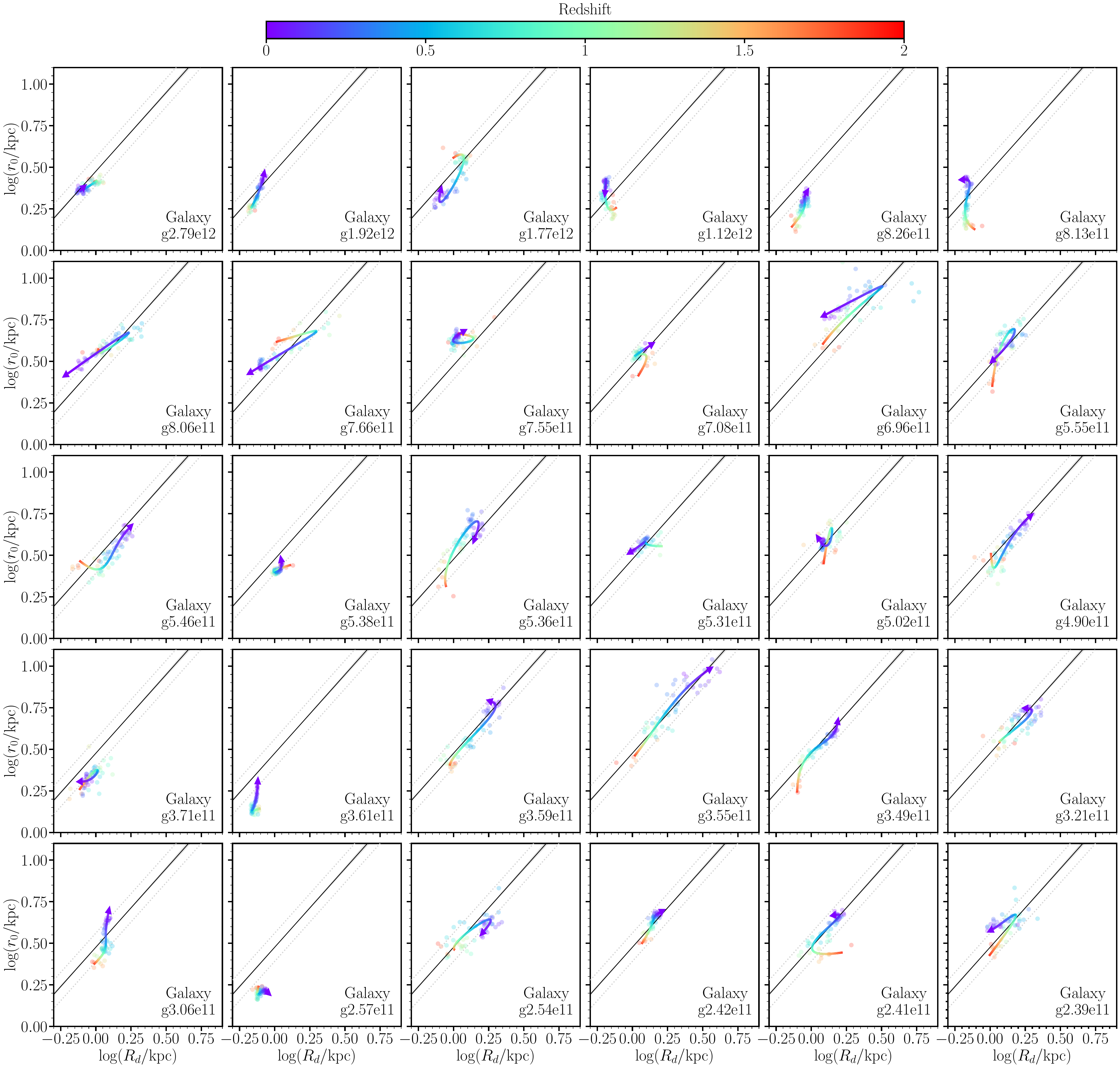}
    \caption{Evolution of the $R_d{-}r_0$ relation for individual galaxies in the NIHAO sample.
Each panel shows the evolutionary track of a single galaxy in the $\log(R_d)$--$\log(r_0)$ plane, color-coded by redshift from $z=2$ (red) to $z=0$ (violet).
The trajectories have been smoothed using spline interpolation, and arrows indicate the direction of evolution toward decreasing redshift.
The black solid curve represents the median model relation at $z=0$, while the black shaded region marks its $1\sigma$ credible interval derived from the posterior distribution (see \autoref{time_evolution}).
The light-gray dotted lines indicate the intrinsic scatter of $\sim0.085$ dex for the relation at $z=0$.
Galaxies are ordered by decreasing stellar mass from the top-left to the bottom-right panels, illustrating how systems of different masses follow distinct evolutionary paths while converging toward the same present-day scaling relation.\\
\emph{Note $-$} To smooth the data points for each galaxy, we adopted a cubic B-spline with smoothing factor $s=1$.}
    \label{fig:all_galaxies}
\end{figure*}

To study the evolution of a single galaxy in the $R_d{-}r_0$ plane, we fit the dark matter and stellar surface density profiles as function of time from \(z = 2\) down to \(z = 0\), Fig. \ref{fig:all_galaxies} shows the trajectory of our galaxies as a function of time in this plane\footnote{In NIHAO the name of a galaxy represents its total mass in the low resolution N-body simulation.}.

It is quite interesting that practically all galaxies move {\it along} the relation (even if in opposite directions), showing a strong correlation in the evolution of these two parameters. 

 Two distinct evolutionary trends emerge from the analysis of the tracks: a subset of galaxies evolves along the $R_d{-}r_0$ relation from left to right, indicating a simultaneous increase in both $R_d$ and $r_0$ over cosmic time. Conversely, another group follows the opposite trajectory, migrating from right to left, suggesting a progressive contraction in both scales. Additionally, a small number of galaxies exhibit more stochastic behavior, scattering around a localized region of the plane without adhering to a clear evolutionary path.

To better quantify the evolutionary behavior of galaxies on the $R_d{-}r_0$ plane, we classify the sample into three distinct groups based on the direction of their evolution. For each galaxy, we compute the change in both $R_d$ and $r_0$ across successive snapshots between \(z = 2\) and \(z = 0\), using a constant time interval of \(\Delta t = 0.23\,\mathrm{Gyr}\), uniformly spaced in cosmic time. \\
Let \(r_0^{(i)}\) and \(R_d^{(i)}\) denote the DM  scale radius and disc scale length at time step \(i\), respectively. The incremental changes between two consecutive steps are given by:
$$
\Delta \log r_0^{(i)} = \log r_0^{(i+1)} - \log r_0^{(i)}, \quad
\Delta \log R_d^{(i)} = \log R_d^{(i+1)} - \log R_d^{(i)}.
$$
The net evolutionary direction of each galaxy is then assessed by averaging over all \(N\) steps:
$$
\langle \Delta \log r_0 \rangle = \frac{1}{N} \sum_{i=1}^{N} \Delta \log r_0^{(i)}, \quad
\langle \Delta \log R_d \rangle = \frac{1}{N} \sum_{i=1}^{N} \Delta \log R_d^{(i)}.
$$

\begin{figure}
    \includegraphics[width=\columnwidth]{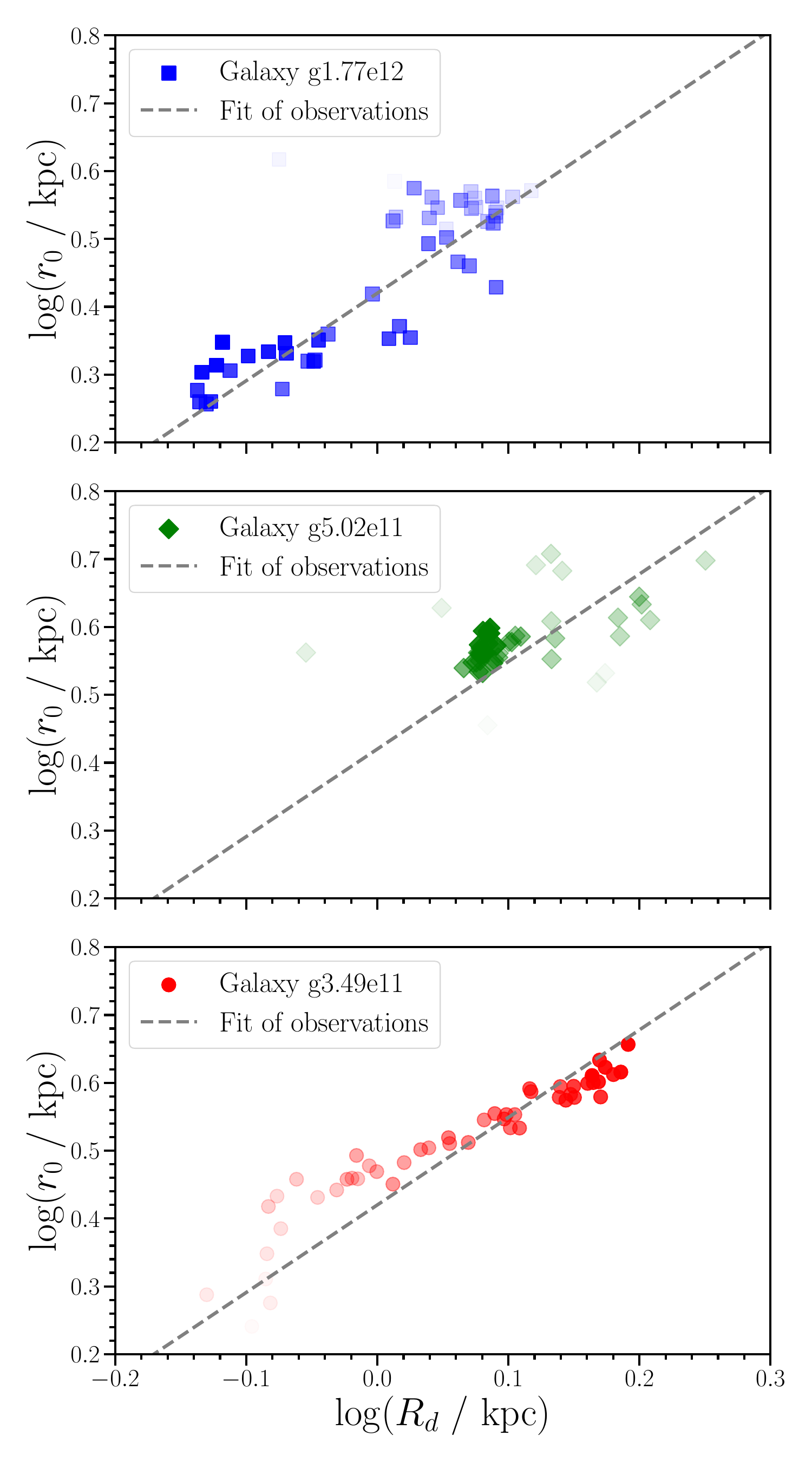}
    \caption{The time evolution of three halos on the $R_d{-}r_0$ relation from $z=2$ to $z=0$ where a fainter point indicates an earlier time. The plot shows an expansion of the $3.49 \times 10^{11} M_\odot$ galaxy and a contraction of the $1.77 \times 10^{12} M_\odot$ galaxy over time, while the third galaxy of mass $5.02 \times 10^{11} M_\odot$ doesn't follow either trend but remains scattered around the same region.}
    \label{fig:3galaxies}
\end{figure}

\begin{figure}
    \includegraphics[width=\columnwidth]{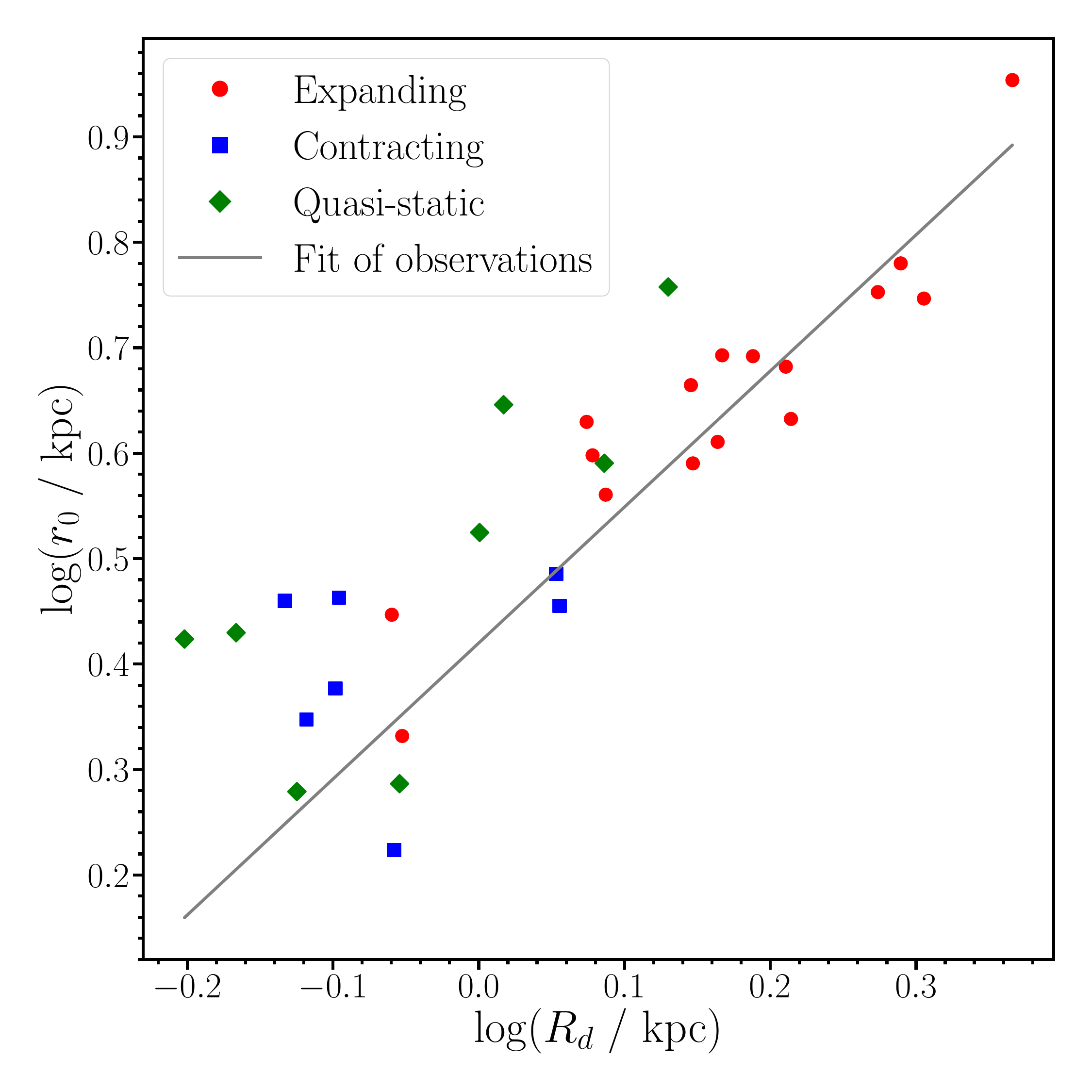}
    \caption{$r_0$ versus $R_d$ for simulated galaxies at $z=0$, divided into three groups, expanding (red dots), contracting (blue squares), and quasi-static galaxies (green diamonds), according to their evolutionary tracks on the plane. The gray line represents the fit derived in this work for observed galaxies. 
    }
    \label{sim_relation}
\end{figure}

Galaxies for which both averages are positive are classified as \textbf{"expanding"} (evolving from left to right); for these galaxies, the dark matter inner core and the stellar disk are growing over time. Those with both averages negative are \textbf{"contracting"} (evolving from right to left), they have both radii decreasing over time, concentrating more mass in the center of the galaxy. Galaxies with mixed or near-zero values are categorized as \textbf{"quasi-static"}, indicating no clear directional trend. Fig. \ref{fig:3galaxies} presents three representative examples, one for each galaxy family, showing their detailed evolutionary histories. Fig. \ref{sim_relation}, instead, displays the full sample of galaxies on the $R_d{-}r_0$ plane, where each galaxy is shown at its final redshift only.

\subsection{The Role of Feedback-driven Interaction}
A first indication that baryonic processes play a key role in shaping the evolution on the $R_d{-}r_0$ plane comes from comparing hydrodynamical simulations with their dark-matter-only (DMO) counterparts. In this section, we adopt an alternative definition of $r_0$ that is better suited to the cuspy density profiles of the DMO runs \citep{NFW1997,maccio2007concentration}. For consistency, the same definition is also applied to the hydrodynamical simulations to enable a direct comparison. We u
adopt the functional form used by \cite{tollet2016nihao}, given by

\begin{equation}
\rho(r) = \rho_0 \exp \left\{ \lambda \left[\ln\!\left(1 + \frac{r}{r_0}\right)\right]^2 \right\}.
\end{equation}

Here, $\rho_0$ denotes the central density, corresponding to the asymptotic value approached as $r \rightarrow 0$. The parameter $r_0$ defines the characteristic radial scale of the profile; we adopt the fitted value of $r_0$ as a measure of the core size. The parameter $\lambda$ controls the curvature of the profile and determines how rapidly the density transitions from the inner region to the outer slope. This functional form is sufficiently flexible to describe both cuspy and cored density profiles.

\begin{figure}
    \includegraphics[width=\columnwidth]{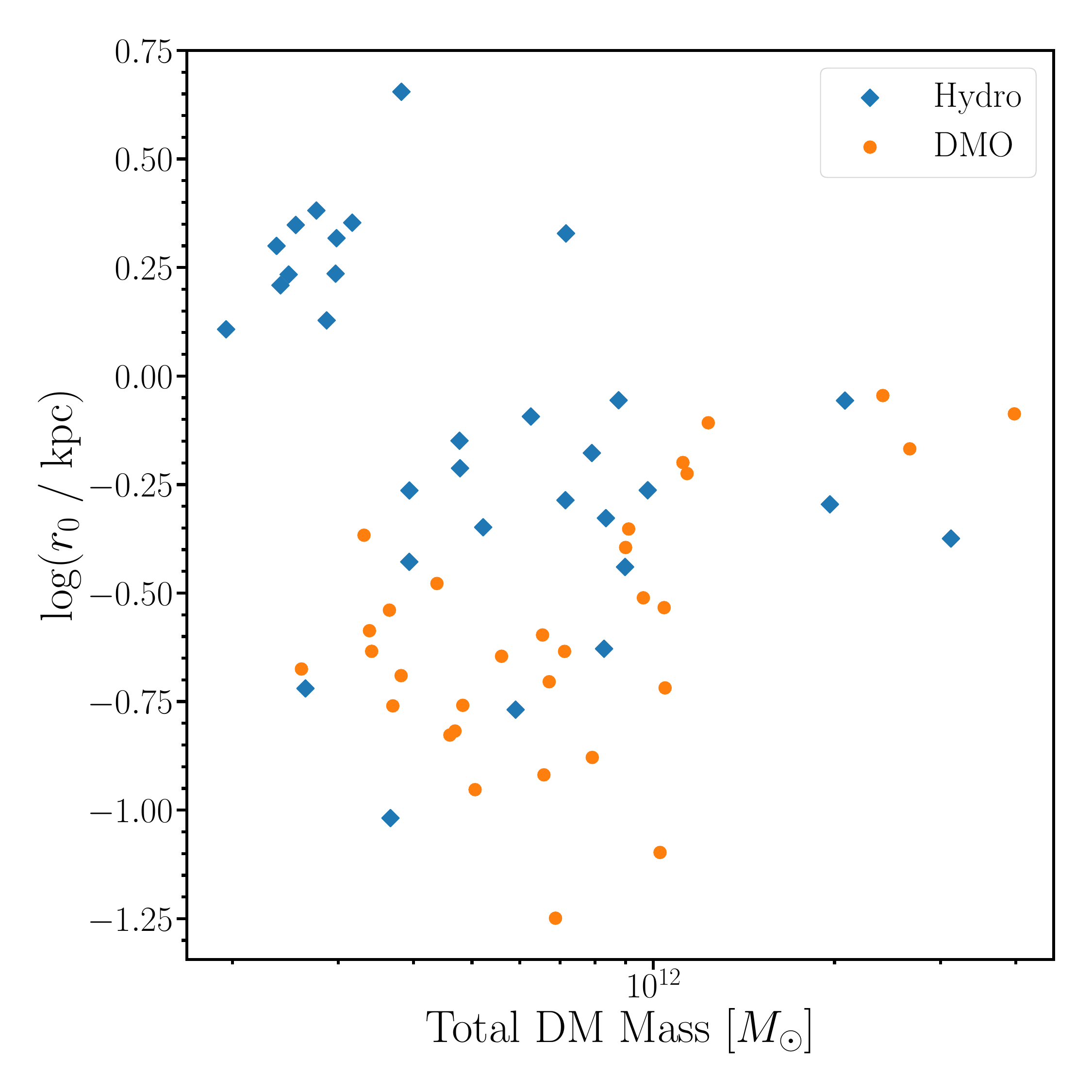}
    \caption{The dark matter core radius $r_0$ versus the total dark matter halo mass of all galaxies at $z=0$.}
    \label{dmo1}
\end{figure}

\begin{figure}
    \includegraphics[width=\columnwidth]{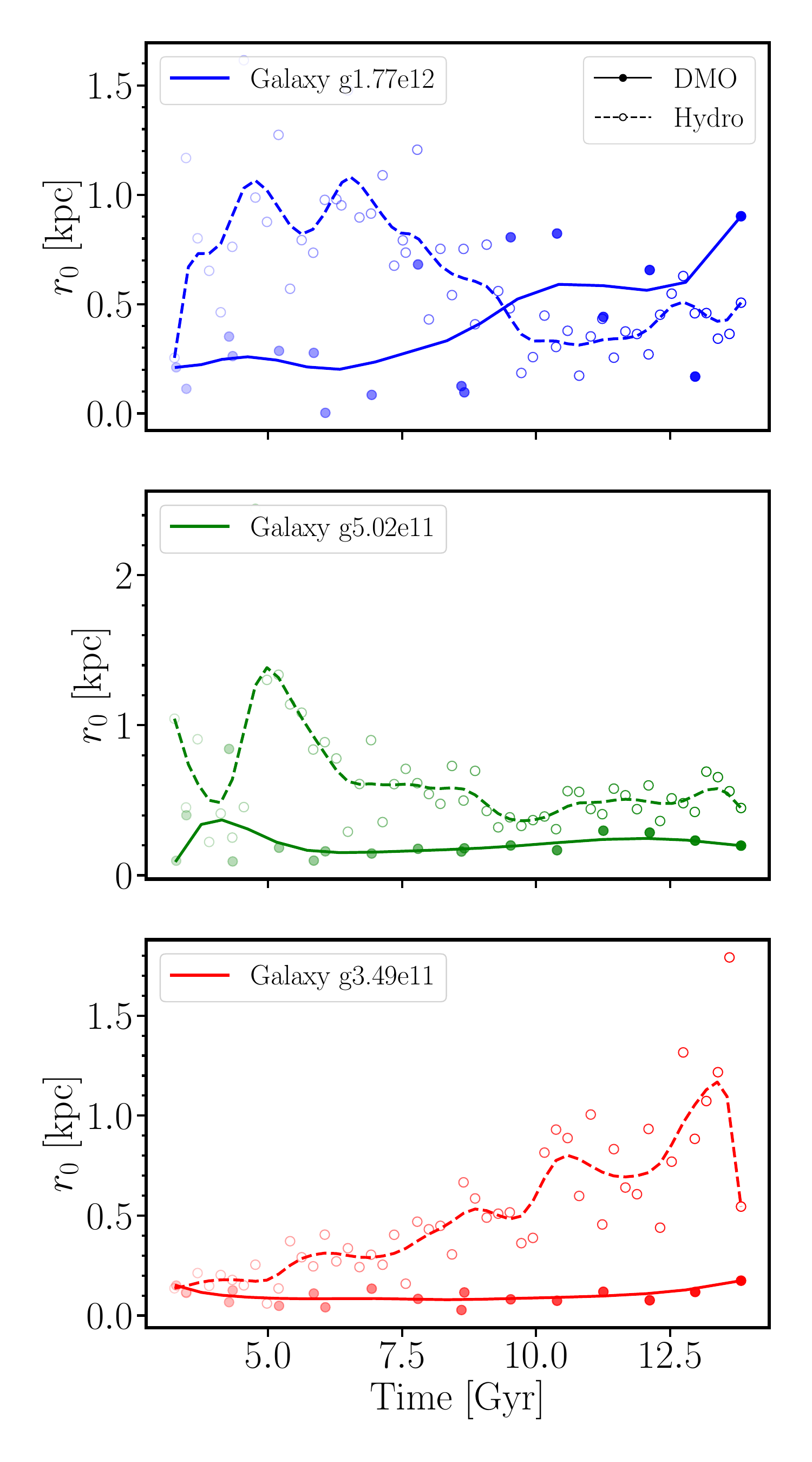}
    \caption{The time evolution of three halos on the $r_0$–dark matter halo mass plane from $z=2$ to $z=0$ comparing Hydro and DMO runs. The figure shows three representative cases: a contracting galaxy with halo mass $1.77 \times 10^{12},M_\odot$ (upper panel); a quasi-static galaxy with mass $5.02 \times 10^{11},M_\odot$ (mid panel); an expanding galaxy with mass $3.49 \times 10^{11},M_\odot$ (lower panel).}
    \label{dmo2}
\end{figure}

Fig. \ref{dmo1} shows the relation between the dark matter core scale radius $r_0$ and halo mass $M_{\rm DM}$ for the NIHAO simulations in both Hydro and DMO runs, while Fig. \ref{dmo2} presents the evolutionary tracks of same three representative galaxies shown before.

In the DMO simulations, the evolution follows smooth and monotonic trends, with $r_0$ varying gradually as halo mass increases through hierarchical assembly. This behavior reflects the purely gravitational evolution of collisionless dark matter. In contrast, the Hydro simulations exhibit markedly different trajectories and a larger scatter, indicating that baryonic processes significantly alter the inner dark matter structure over time.

For the most massive galaxy (g1.77e12, top row), $r_0$ initially increases, indicating halo expansion at early times when the system is still assembling and the gravitational potential is relatively shallow. At later times, as the halo grows in mass and a substantial stellar component builds up at the centre, the potential deepens and the dark matter distribution contracts. As a result, $r_0$ decreases and eventually reaches values smaller than those found in the corresponding DMO run.

The lowest mass galaxy (g3.49e11, low row) shows a markedly different behaviour. In the DMO case, $r_0$ again remains approximately constant, whereas in the hydrodynamical simulation we observe a clear and sustained expansion of the dark matter distribution. Repeated gas outflows driven by supernova feedback significantly perturb the gravitational potential, forcing the dark matter to respond dynamically by migrating to larger radii \citep{Pontzen2012, tollet2016nihao}. This galaxy a clear indication of feedback-driven baryon--dark matter coupling. By the end of the simulation, $r_0$ has increased by roughly a factor of $5$--$10$ compared to its initial value.

Finally, the intermediate-mass galaxy (g5.02e11, central row) displays a behaviour in between these two extremes. The hydrodynamical run yields systematically larger values of $r_0$ than in the DMO case, indicating moderate halo expansion. However, aside from the early phase of halo assembly, $r_0$ remains approximately constant with time, suggesting that neither strong contraction nor sustained expansion dominates the subsequent evolution.

The NIHAO DMO runs therefore provide a useful baseline for purely gravitational evolution, allowing the deviations seen in the hydrodynamical simulations to be interpreted as the result of baryons an baryonic  feedback acting on the dark matter distribution.

\begin{figure}
    \includegraphics[width=\columnwidth]{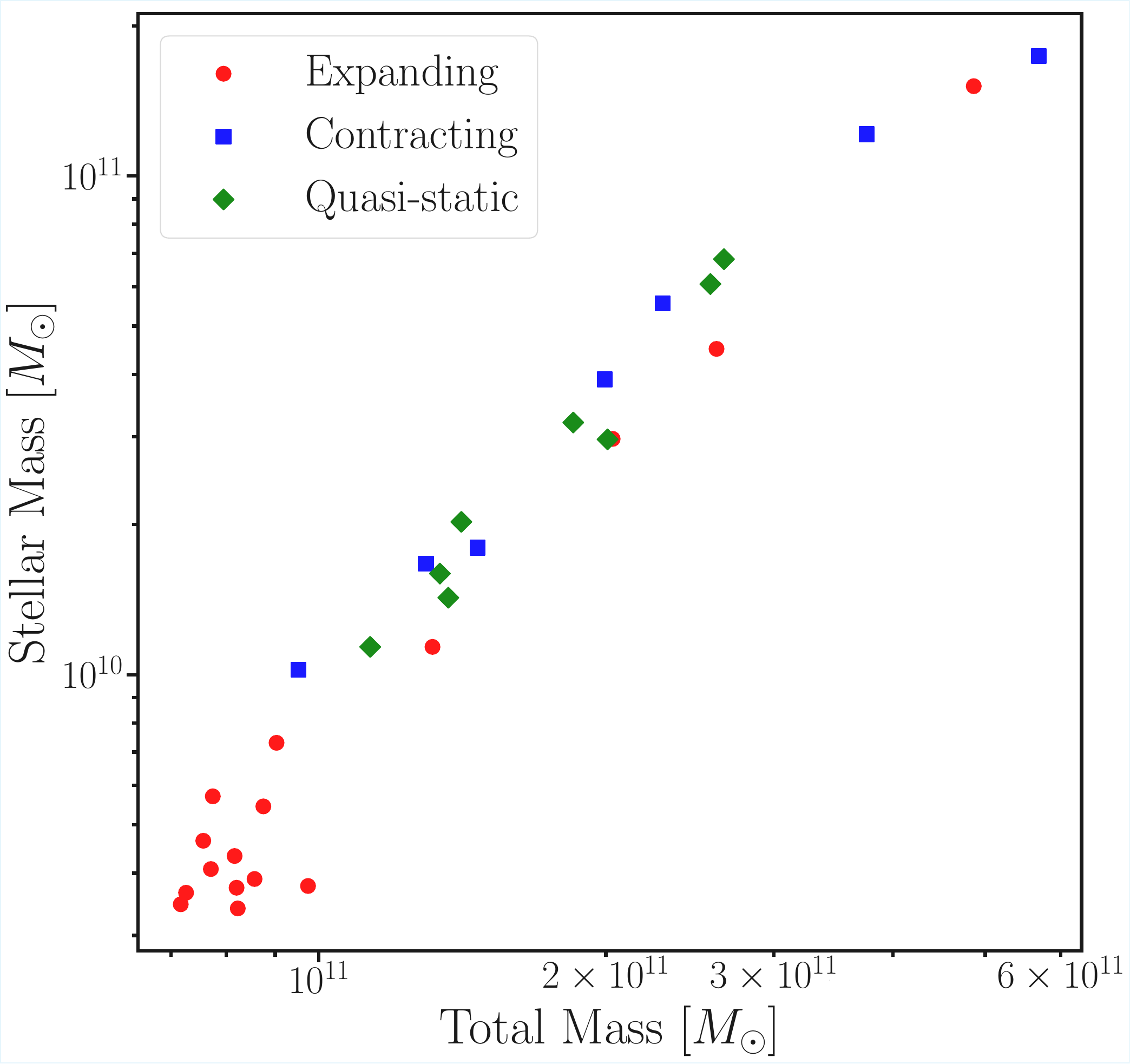}
    \caption{The stellar mass versus the total halo mass of all galaxies at $z=0$.}
    \label{mass}
\end{figure}

To better understand the origin of the different evolutionary behaviors observed on the $R_d{-}r_0$ plane, we examined a range of physical properties across the three classified groups. Our analysis indicates that stellar and halo mass are key drivers of the evolutionary direction, as shown in Fig. \ref{mass}. Specifically, galaxies with total masses (\(M_{\mathrm{tot}}\)) lower than $10^{11} M_{\odot}$ dominantly exhibit outward evolution, characterized by a monotonic increase in both \(r_0\) and \(R_d\) over time. In contrast, more massive systems tend to follow an inward evolutionary path, with both characteristic radii decreasing across cosmic time. This mass-dependent behavior suggests that shallower potential wells in low-mass halos allow for sustained expansion of baryonic and dark matter distributions, possibly driven by feedback processes \citep[e.g.][]{Pontzen2012, tollet2016nihao}, while at higher masses the additional potential due to the presence of stellar bulge forces the dark matter to contract \citep[e.g.][]{abadi2010,maccio2020nihao}.

\begin{figure}
    \includegraphics[width=\columnwidth]{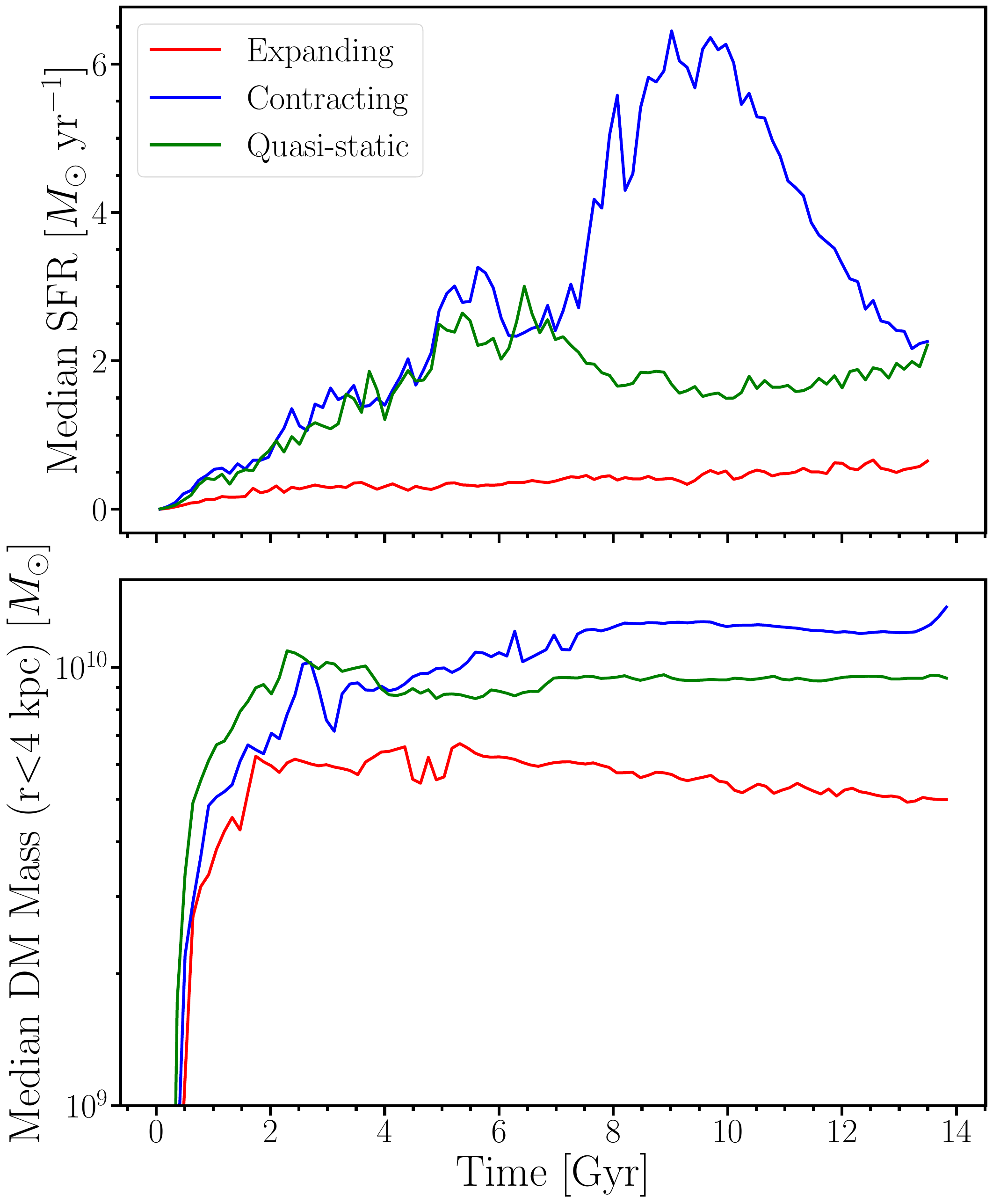}
    \caption{The top plot shows the median star formation rate over galaxies in each group according to how the galaxy evolves on the $R_d{-}r_0$ relation, while the bottom plot shows the median dark matter mass within 4 kpc for each group.}
    \label{SFR}
    \label{dm_mass}
    \label{dm_rate}
\end{figure}

This interpretation is confirmed by the strong correlation of the direction of the movement in the $R_d{-}r_0$ plane and the star formation rate (SFR) of the galaxy (as shown in Fig. \ref{SFR}).   Galaxies with low or declining SFR tend to show expanding behavior, consistent with a scenario of gradual disc growth and DM core creation \citep{tollet2016nihao}, while galaxies exhibiting stronger star formation rates (which are usually associated with bulge formation \citep[e.g.][]{Brook2011}) are more likely to fall into the contracting category.

The ability of baryons to influence and regulate the amount of dark matter in the central region of a galaxy is clearly visible in Fig. \ref{dm_mass} where we show the time evolution of the DM mass within a comoving radius of 4 kpc for our three track families.  One more time, it is possible to appreciate the {\emph{interaction} between visible and invisible matter even in the $\Lambda$CDM framework, which does not include any direct interaction mechanism.

Finally, not surprisingly, we find a strong correlation between the inner slope of the dark matter halo profile measured within $2\%$ of the virial radius \citep[][]{tollet2016nihao} and the trajectory in the $R_d{-}r_0$ plane, as shown in Fig. \ref{dm_rate}. It is interesting to note that quasi-static galaxies fall in the region where dark matter haloes have retained the original NFW-like profile \citep{navarro2000dark}, showing no net evolution in their dark matter distribution.

\begin{figure}
    \includegraphics[width=\columnwidth]{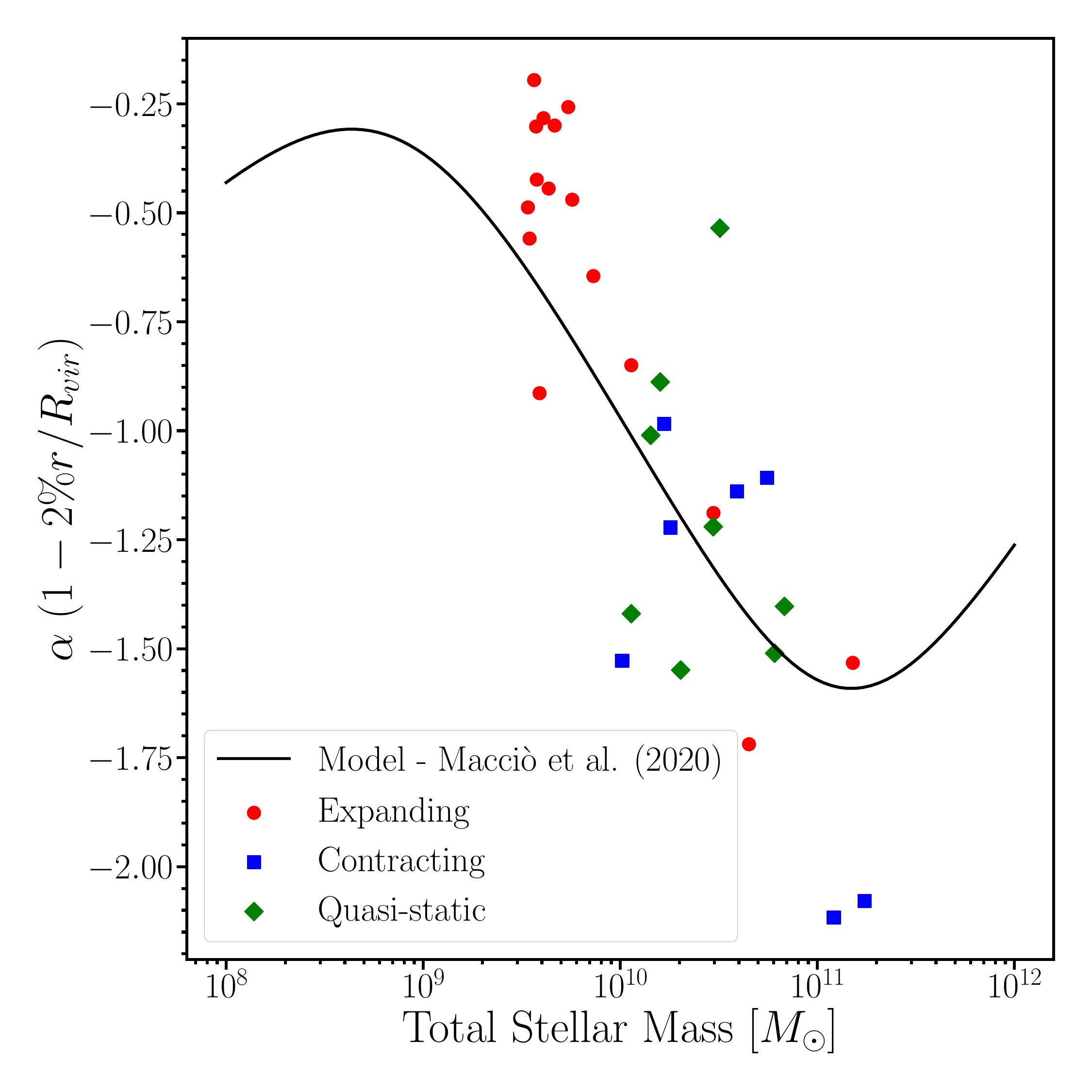}
    \caption{Dark matter halo inner density profile slope $\alpha$ as a function of the stellar mass at redshift $z = 0$. The model is retrieved from \citet{maccio2020nihao}.}
    \label{alpha_mstar}
\end{figure}

All of the above results seem to strongly suggest a dark-baryonic matter \emph{interaction} via feedback mechanisms. The gentle build-up of the stellar disc (with moderate SFR) is accompanied by stellar feedback, which is able to displace gas from the center of the galaxy in the lower mass haloes in our sample. This gas sloshing generates a time variable potential which induces a non-adiabatic expansion in the DM halo, as initially described in \cite{Pontzen2012}. The net effect is that both the disc and dark matter scale radius grow as a function of time. In massive haloes, vigorous SF in the central region causes the build-up of stellar bulge, which has the of reducing the disc scale radius and contracting the dark matter towards the center. The net effect is that both $R_d$ and $r_0$ decrease as a function of time (contraction).

Both \emph{expansions} and \emph{contractions} operate to varying degrees in all systems, but it is the dominance of one mechanism over the other that dictates the long-term evolutionary path of the galaxy.

\section{The redshift evolution of the ${R_{\text{\lowercase{d}}}}$-$\text{\lowercase{r}}_0$ relation} 

\label{time_evolution}

Given the ability of cosmological simulations to trace galaxy evolution across cosmic time, we utilize the simulation outputs at various redshifts to extract the values of $r_0$ and $R_d$, thereby reconstructing the temporal evolution of the $R_d{-}r_0$ relation.

\subsection{The model}
As already noted in \autoref{tracks}, galaxies evolve by gradually changing their physical properties, including both $r_0$ and $R_d$, eventually approaching the linear regime observed in the present-day Universe.
We now aim to provide a quantitative description of how the global relation between $r_0$ and $R_d$ evolves continuously throughout cosmic history, analyzing how its intrinsic scatter varies as well. For this scope, similarly to the approach adopted in \citet{Waterval2025MNRAS} (see also \citealt{Cannarozzo2020MNRAS}), we make use of Bayesian hierarchical modeling to reconstruct the evolution of the relation between $r_0$ and $R_d$.


We adopt a single power-law evolutionary model in which the quantity $\log r_0$ depends linearly on $\log R_d$ with coefficients evolving with redshift. The relation is expressed as
\begin{equation}
\mu(z, \log R_d) \;=\; \log r_{0}(z) + \alpha(z)\,\log R_d,
\label{eq:mean}
\end{equation}
with
\begin{equation}
    \log r_{0}(z) = \log r_{0,0} + \log r_{0,z}\,z,
\label{eq:normalization}
\end{equation}
 and
\begin{equation}
    \alpha(z) = \alpha_0 + \alpha_z\,z.   
\label{eq:slope}
\end{equation}

Furthermore, to account for intrinsic scatter and any possible evolution of that, we assume the data follow a Gaussian distribution centered on $\mu$ with variance $\sigma^2(z)$, i.e.
\begin{equation}
\mathcal{P}(\log R_0 \,|\, \boldsymbol{\Theta}) \;=\; \mathcal{N}\!\big(\mu(z, \log R_d),\, \sigma^2(z)\big),
\label{eq:probability}
\end{equation}
where $\boldsymbol{\Theta} = \{\log r_{0,0}; \log r_{0,z}; \alpha_0; \alpha_z; \sigma_0; \sigma_z\}$ is the set of parameters to describe the evolutionary single power-law relation.
The standard deviation in \autoref{eq:probability} is modeled as
\begin{equation}
    \sigma(z) = \sigma_0 + \sigma_z z.
\label{eq:scatter}
\end{equation}
The corresponding log-likelihood is
\begin{equation}
\ln \mathcal{L} = -\tfrac{1}{2}\sum_i \left( \frac{(\log r_{0,i} - \mu_i)^2}{\sigma_i^2} + \ln\!\big(2\pi\sigma_i^2\big) \right).
\end{equation}
where the index $i$ runs over the individual NIHAO galaxies.

Further details about the parameters adopted and their prior ranges, as well as their posterior, are presented in Appendix \ref{app:model}.

\subsection{The results}

We performed the parameter inference using the \emph{Nested Sampling} algorithm, as implemented in the \textsc{dynesty} package \citep{Speagle2020MNRAS,Koposov2023zndo}. The sampling setup and the adopted priors for the model hyperparameters are described in detail in Appendix~\ref{app:model}.
The best-fitting relations at different redshifts are shown in \autoref{fig:model}, color-coded by $z$. The model predicts a gradual steepening of the $\log r_0$–$\log R_d$ relation toward higher redshift, while simultaneously reproducing the progressive tightening of its intrinsic scatter at later times. The median values and 68\% credible intervals of all hyperparameters are listed in \autoref{tab:posteriors}.

At $z=0$, our best-fit relation (violet curve in \autoref{fig:model}) lies slightly below the local calibration by \citet{nesti2023quest}, shown as a gray dashed line. This offset originates from differences in the inferred slopes: while \citet{nesti2023quest} report $\alpha = 1.38 \pm 0.15$ and $\log r_{0,0} = 0.47 \pm 0.03$, our posterior analysis yields $\alpha(z=0) \simeq 0.95$ and $\log r_{0,0}(z=0) \simeq 0.48$ (see \autoref{fig:posteriors}). The two results are therefore consistent in normalization but differ significantly in slope, implying a shallower dependence of $\log r_0$ on $\log R_d$ in our model. Quantitatively, our relation lies $\sim+0.08$~dex above the \citet{nesti2023quest} relation at $\log R_d = -0.2,\kpc$, becomes nearly coincident around $\log R_d \simeq 0$, and falls progressively below it by $-0.08$ and $-0.15$~dex at $\log R_d = 0.2$ and $0.4$, respectively. This systematic deviation, clearly visible in \autoref{fig:model}, reflects the flatter slope inferred from our nested-sampling posterior, despite the overall agreement in normalization.

We find that, from $z=2$ to $z=0$, the normalization decreases by $\sim0.07$ dex, while the slope flattens from $\alpha\simeq1.05$ to $\alpha\simeq0.95$, indicating a mild but coherent evolution toward shallower trends at later times. This behaviour reflects the gradual structural coevolution of the stellar and dark matter components, which become more tightly coupled as the Universe ages.

The inset panel in \autoref{fig:model} illustrates the weak redshift evolution of the intrinsic scatter $\sigma(z)$, which increases from $\sigma \simeq 0.086$~dex at $z=0$ to $\sigma \simeq 0.094$~dex at $z=2$. This slight broadening indicates that, although the $R_d{-}r_0$ relation is already tight at late times, it becomes marginally more dispersed at earlier epochs. Such an evolution suggests that the structural diversity of galaxies was larger when disks were still assembling and baryon–dark matter coupling was more dynamically variable. At high redshift, stochastic processes—such as bursty star formation, feedback-driven potential fluctuations, and irregular accretion histories—likely contributed to this increased dispersion. By contrast, toward $z=0$, the reduced scatter reflects the secular stabilization of disks and the emergence of a self-regulated equilibrium between stellar and dark matter scales.

In summary, the evolution of the $R_d{-}r_0$ relation across the redshift range $0 \lesssim z \lesssim 2$ can be approximately described as
\begin{equation}
\log \left(\frac{r_0}{\kpc}\right) \simeq (0.476 - 0.036z) + (0.946 + 0.104z)\log R_d,
\end{equation}
based on the median posterior values of the hyperparameters.
The weak evolution of the intrinsic scatter is captured by
\begin{equation}
\log \left(\frac{\sigma}{\mathrm{dex}}\right) \simeq 0.086 + 0.004z,
\end{equation}
indicating a mild increase of scatter with redshift, consistent with the trend shown in the inset of \autoref{fig:model}.

\begin{figure}
     \includegraphics[width=1\columnwidth]{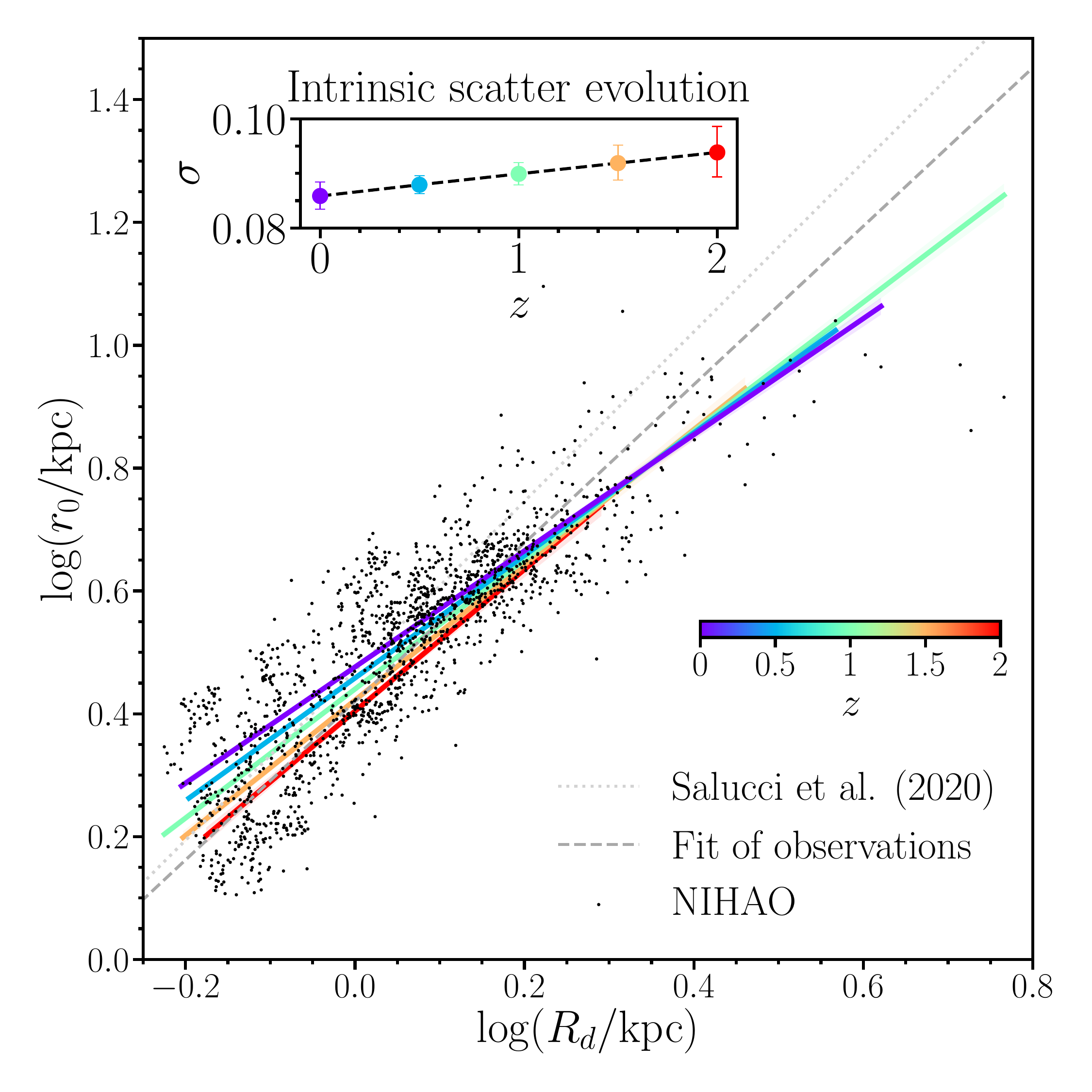}
     \caption{Cosmic evolution of the $R_d{-}r_0$ relation from $z=2$ to $z=0$. The solid colored lines show the median relation at different redshifts ($z = 0,\ 0.5,\ 1,\ 1.5,\ 2$). The shaded bands represent the $1\sigma$ credible regions from the Bayesian hierarchical model. The black points correspond to all NIHAO galaxies across all available snapshots between $z=2$ and $z=0$. The light gray dotted line shows the relation from \citet{salucci2020paradigms}, while the dark gray dashed line represents the linear fit to the full compilation of observational data presented in \autoref{obs}. The inset panel illustrates the redshift evolution of the intrinsic scatter (black dashed line), with the colored points and error bars indicating the median and 16th–84th percentile ranges of the posterior distributions at each redshift.}
     \label{fig:model}
 \end{figure}

\begin{table}
\centering
\caption{Median values and 68\% credible intervals of the posterior distributions for the hyperparameters of the evolutionary model.}
\renewcommand{\arraystretch}{1.3} 
\setlength{\tabcolsep}{8pt}      
\begin{tabular}{lc}
\hline
Hyperparameter & Inferred value \\
\hline
$\log r_{0,0}$ &  $\displaystyle {0.476}^{+0.004}_{-0.004}$ \\[0.2em]
$\log r_{0,z}$ & $\displaystyle {-0.036}^{+0.005}_{-0.005}$ \\[0.2em]
$\alpha_0$     &  $\displaystyle {0.946}^{+0.024}_{-0.023}$ \\[0.2em]
$\alpha_z$     &  $\displaystyle {0.104}^{+0.037}_{-0.037}$ \\[0.2em]
$\sigma_0$     &  $\displaystyle {0.086}^{+0.003}_{-0.002}$ \\[0.2em]
$\sigma_z$     &  $\displaystyle {0.004}^{+0.003}_{-0.003}$ \\[0.2em]
\hline
\end{tabular}
\label{tab:posteriors}
\end{table}

\begin{figure}
     \includegraphics[width=1\columnwidth]{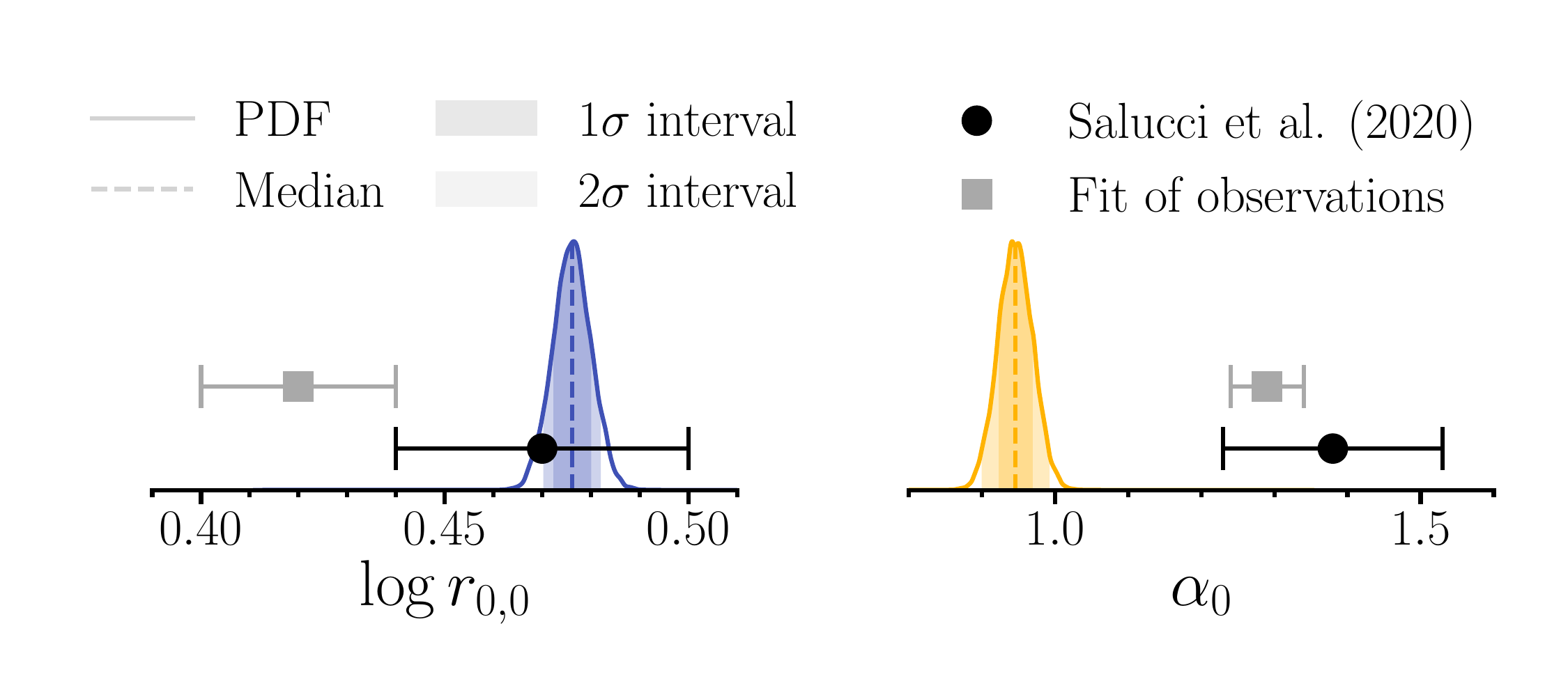}
     \caption{Posterior probability density functions (PDFs) for the normalization ($\log r_{0,0}$, left) and slope ($\alpha_0$, right) of the $R_d{-}r_0$ relation at $z=0$. The colored curves show the marginalized PDFs, with dashed vertical lines marking the median values and shaded regions indicating the $1\sigma$ (dark) and $2\sigma$ (light) credible intervals. Black circles and horizontal error bars denote the observational estimates from \citet{salucci2020paradigms}, while gray squares represent the best-fit values obtained from the linear fit to the full set of observational data compiled from the literature, as described in \autoref{obs}. The comparison highlights the close agreement between the simulated posterior distributions and the empirical measurements at $z=0$.}
     \label{fig:posteriors}
 \end{figure}


In \autoref{fig:cartoon}, we show an interpretative cartoon that illustrates how galaxies follow different evolutionary pathways in the relation between the disk scale length and the DM scale radius, before all converging onto the observed tight relation at $z=0$. Expanding systems (red) increase both $R_d$ and $r_0$ with time, while contracting systems (blue) decrease both quantities, and “quasi-static” galaxies (\textcolor{red}{green}) remain relatively stable, only oscillating around the final trend. These structural differences are closely connected to their star formation histories: expanding galaxies maintain a low and flat star formation rate, contracting galaxies sustain a high and extended star formation activity, and “quasi-static” galaxies show a pronounced early peak followed by a steady decline. The inset with concentric circles further emphasizes how the two characteristic radii evolve in each category, highlighting the distinct evolutionary tracks that nonetheless collapse onto the same relation at the present-day Universe.
 
\begin{figure*}
     \includegraphics[width=1.85\columnwidth]{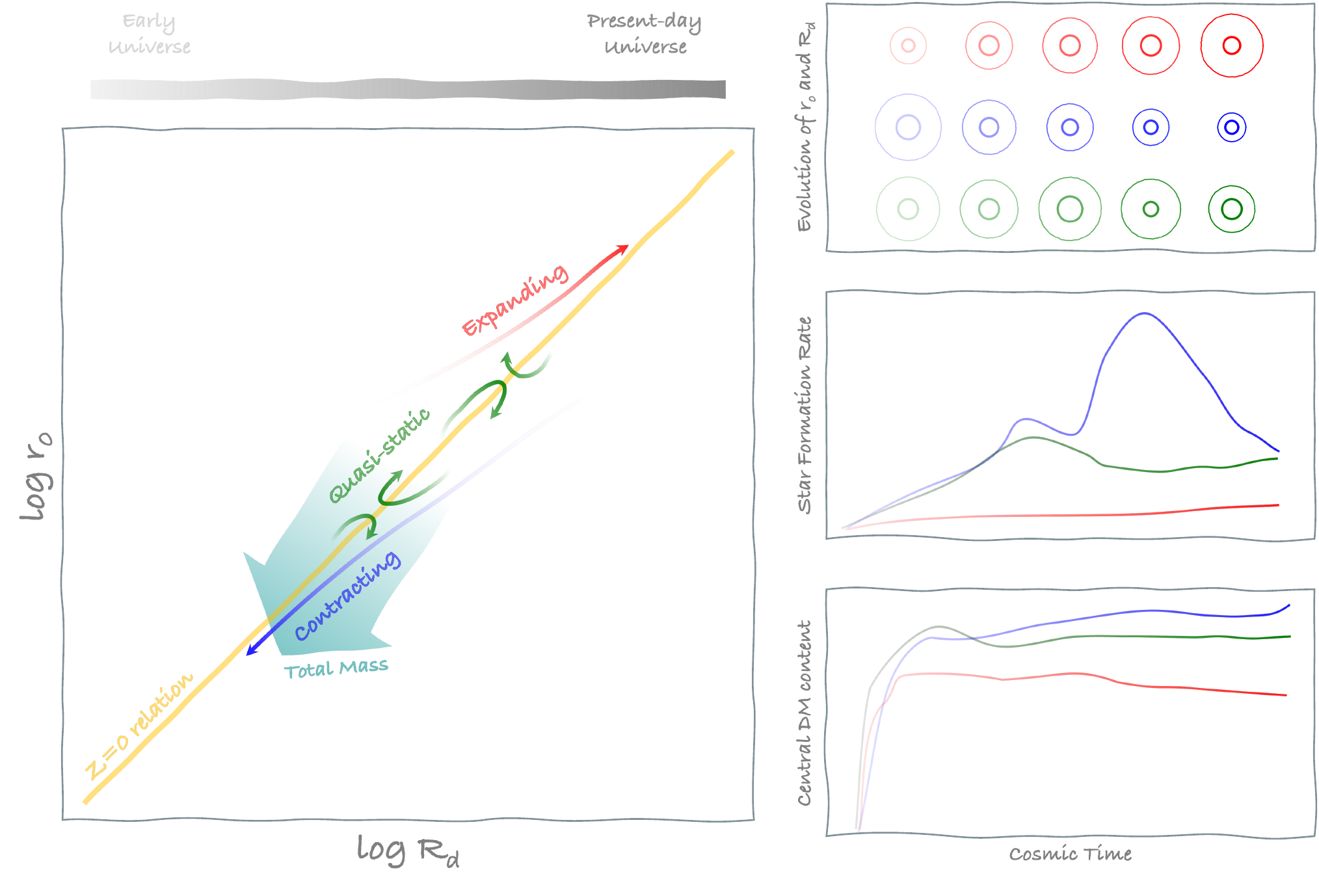}
     \caption{Cartoon representation of the evolutionary pathways of galaxies in the $R_d$–$r_0$ plane and related quantities.
The left panel illustrates the average evolutionary trajectories of “expanding” (red), “contracting” (blue), and “quasi-static” (green) galaxies, which follow distinct pathways but converge toward the same tight $z=0$ relation (orange line).
The large teal arrow indicates the overall trend with galaxy mass at $z=0$.
The top-right panel shows the evolution of $r_0$ (thick circle) and $R_d$ (thin circle), the middle panel the star formation rate (SFR), and the bottom panel the inner dark matter content.
Color gradients along each track, curve, and on the concentric circles (solid for stellar disks, dotted for dark matter scale radius) trace cosmic time, with more intense hues marking more recent epochs.
These panels illustrate how structural and baryonic properties coevolve along different evolutionary channels while maintaining a coherent link between disk growth, central mass buildup, and global size evolution.}
     \label{fig:cartoon}
 \end{figure*}
 


\section{Caveats} \label{caveats}

\begin{itemize}

    \item \emph{Intrinsic differences between observations and simulations.} 
    While both datasets aim to describe the same physical systems, simulations rely on sub-grid physics and resolution limits \citep{hopkins2018fire}, whereas observations are constrained by instrumental sensitivity, projection effects, and selection biases \citep{conselice2014evolution}. These intrinsic differences must be acknowledged when drawing one-to-one comparisons \citep{somerville2015physical, naab2017theoretical}.

    \item \emph{Different methods to derive physical quantities.}
    Stellar masses or radii are often obtained using distinct techniques in observations (e.g., photometric or spectroscopic fitting) and simulations (e.g., particle summation within a given metric) \citep{torrey2015synthetic}. Such methodological inconsistencies can lead to systematic offsets in derived relations.

    \item \emph{Choice of radial profiles and metrics.}
    Observational studies typically measure quantities within projected metrics (e.g., $R_e$, or fixed physical sizes), while simulations can access full 3D profiles \citep{schaye2015eagle}. The choice of metric or profile definition strongly impacts global quantities and scaling relations \citep{pillepich2018first, rodriguez2019optical}. Moreover, the adopted photometric band also affects measured galaxy sizes: multi-band analyses show systematic differences of effective radius \citep[e.g.,][]{vulcani2014wavelength, kennedy2015gama}, and mock radiative-transfer predictions from cosmological simulations (e.g., TNG50–SKIRT) reveal size variations of up to $\sim50\%$ across different bands \citep{baes2024tng50skirt}.
    We emphasize here that, in our analysis, we did not perform mock observations of the simulated galaxies to produce wavelength-dependent maps directly comparable to observations, as such an analysis lies beyond the scope of this work.

    \item \emph{Impact of resolution and numerical effects.}
    Finite spatial and mass resolution in simulations can affect the internal structure of galaxies, especially at low masses, influencing size, morphology, and kinematic measurements \citep{hopkins2018fire,buck2020nihao}. Resolution convergence tests are essential for robust interpretations.

    \item \emph{Uncertainties in stellar population modeling.}
    Translating between observed light and simulated stellar mass requires assumptions about IMF, stellar evolution, and dust attenuation. Differences in these assumptions can introduce additional scatter and bias \citep{macarthur2003structure, conroy2009propagation}.


\end{itemize}

\section{Conclusions} \label{conc}

Galaxy scaling relations are a natural testbed for any models aiming to reproduce the formation and evolution of galaxies through cosmic time.
In this work, we focus our attention on the observed tight correlations between the scale radius of the dark matter distribution, usually obtained via the study of rotation curves \citep{di2019universal} and the exponential scale radius of the stellar disc \citep[e.g.][]{begeman1991extended, karukes2017universal}.

This tight relation between the distribution of the invisible and visible matter might suggest a possible interaction in the dark sector beyond gravity \citep[e.g.][]{nesti2023quest}; it is then important to check to what extent such a relation can be reproduced in the standard $\Lambda$CDM model, which does not include any additional interaction.

We analyzed 31 high-resolution simulated galaxies from the NIHAO suite \citep{wang2015nihao} in a mass range similar to the observed one. 
Our results show that it is indeed possible to reproduce a very tight relation with very low scatter (less than 0.09 dex) in a model where the dark and luminous matter only interact via gravity.

The correlation between the stellar and dark matter distributions is already in place at $z=2$, even though with a slightly larger scatter than at $z=0$. 
Subsequently, galaxies tend to move {\it along} the relation by either expanding or contracting the scale radius of both the distributions.
To quantitatively trace how the $R_d${–}$r_0$ relation evolves across cosmic time, we rely on a Bayesian hierarchical approach. The mild decrease in normalization and the progressive flattening of the slope toward lower redshift indicate that the structural coupling between stellar and dark matter components becomes tighter with time. Galaxies appear to evolve along the $R_d$–$r_0$ relation, maintaining a self-regulated balance between their baryonic and dark matter components as the Universe ages.
Furthermore, we find that the intrinsic scatter weakly decreases toward lower redshift, indicating that galaxies, regardless of their individual evolutionary pathways, tend to evolve coherently along the relation, re-balancing their stellar and dark matter scales.

Galaxies with vigorous star formation histories tend to form a larger bulge, smaller disc, and contract (almost adiabatically) the DM distribution \citep{Blumenthal1986AB, abadi2010}, moving then right to left and top to bottom in the $R_d{-}r_0$ plane.
Galaxies with a more gentle star formation history then to form larger discs, and expand the DM halo via SuperNovae feedback \citep{Pontzen2012, maccio2012ApJ, tollet2016nihao} moving from left to right and bottom to top in the $R_d{-}r_0$ plane.

Our results indicate that \textit{star formation and feedback-driven baryonic processes} can account for the emergence of a tight correlation between the stellar disk exponential radius $R_d$ and the dark matter scale radius $r_0$, without requiring interactions beyond gravity as a natural outcome of galaxy formation.

In sum, the $\Lambda$CDM model—despite its minimal assumptions regarding dark matter interactions—continues to provide a compelling and predictive framework for understanding the structural regularities of galaxies.

\section*{Acknowledgements}
The authors thank Paolo Salucci for his valuable insights during the early stages of this work. We also thank Stéphane Courteau for his valuable discussions and constructive feedback, which helped improve this work. This material is based on work supported by Tamkeen under the NYU Abu Dhabi Research Institute grant CASS.

\section*{Data Availability}

The data underlying this article will be shared on reasonable request
to the corresponding author.



\bibliographystyle{mnras}
\bibliography{example} 

@ARTICLE{NFW1997,
       author = {{Navarro}, Julio F. and {Frenk}, Carlos S. and {White}, Simon D.~M.},
        title = "{A Universal Density Profile from Hierarchical Clustering}",
      journal = {\apj},
         year = 1997,
        month = dec,
       volume = {490},
       number = {2},
        pages = {493-508},
          doi = {10.1086/304888},
archivePrefix = {arXiv},
       eprint = {astro-ph/9611107},
 primaryClass = {astro-ph},
       adsurl = {https://ui.adsabs.harvard.edu/abs/1997ApJ...490..493N}
}

@article{kennedy2015gama,
  author       = {Kennedy, Rebecca and Bamford, Steven P. and Häußler, Boris and Brough, Sarah and Driver, Simon P. and others},
  title        = {Galaxy And Mass Assembly (GAMA): the wavelength dependence of galaxy structure—implications for morphological classification},
  journal      = {Monthly Notices of the Royal Astronomical Society},
  volume       = {454},
  number       = {1},
  pages        = {806--826},
  year         = {2015},
  doi          = {10.1093/mnras/stv2033}
}

@article{vulcani2014wavelength,
  author       = {Vulcani, Benedetta and Bamford, Steven P. and Häußler, Boris and Vika, Marina and Brough, Sarah and others},
  title        = {Galaxy And Mass Assembly (GAMA): the wavelength dependence of galaxy structure—stellar mass and S{\'e}rsic index},
  journal      = {The Astrophysical Journal},
  volume       = {788},
  number       = {1},
  pages        = {57},
  year         = {2014},
  doi          = {10.1088/0004-637X/788/1/57}
}

@article{baes2024tng50skirt,
  author       = {Baes, Maarten and Camps, Peter and Rodriguez-Gomez, Vicente and Nelson, Dylan and Pillepich, Annalisa and others},
  title        = {TNG50-SKIRT: The multi-wavelength mock observables of the IllustrisTNG50 simulation},
  journal      = {Astronomy \& Astrophysics},
  volume       = {690},
  pages        = {A40},
  year         = {2024},
  doi          = {10.1051/0004-6361/202348619}
}

@article{wang2015nihao,
  title={NIHAO project--I. Reproducing the inefficiency of galaxy formation across cosmic time with a large sample of cosmological hydrodynamical simulations},
  author={Wang, Liang and Dutton, Aaron A and Stinson, Gregory S and Macci{\`o}, Andrea V and Penzo, Camilla and Kang, Xi and Keller, Ben W and Wadsley, James},
  journal={Monthly Notices of the Royal Astronomical Society},
  volume={454},
  number={1},
  pages={83--94},
  year={2015},
  publisher={Oxford University Press}
}

@article{blank2019nihao,
  title={NIHAO--XXII. Introducing black hole formation, accretion, and feedback into the NIHAO simulation suite},
  author={Blank, Marvin and Macci{\`o}, Andrea V and Dutton, Aaron A and Obreja, Aura},
  journal={Monthly Notices of the Royal Astronomical Society},
  volume={487},
  number={4},
  pages={5476--5489},
  year={2019},
  publisher={Oxford University Press}
}

@article{wadsley2017gasoline2,
  title={Gasoline2: a modern smoothed particle hydrodynamics code},
  author={Wadsley, James W and Keller, Benjamin W and Quinn, Thomas R},
  journal={Monthly Notices of the Royal Astronomical Society},
  volume={471},
  number={2},
  pages={2357--2369},
  year={2017},
  publisher={Oxford University Press}
}

@article{stinson2013making,
  title={Making Galaxies In a Cosmological Context: the need for early stellar feedback},
  author={Stinson, GS and Brook, C and Macci{\`o}, AV and Wadsley, J and Quinn, TR and Couchman, HMP},
  journal={Monthly Notices of the Royal Astronomical Society},
  volume={428},
  number={1},
  pages={129--140},
  year={2013},
  publisher={Oxford University Press}
}

@article{ade2016planck,
  title={Planck 2015 results-xiii. cosmological parameters},
  author={Ade, Peter AR and Aghanim, Nabila and Arnaud, M and Ashdown, Mark and Aumont, Jea and Baccigalupi, Carlo and Banday, AJ and Barreiro, RB and Bartlett, JG and Bartolo, Nicola and others},
  journal={Astronomy \& Astrophysics},
  volume={594},
  pages={A13},
  year={2016},
  publisher={EDP sciences}
}

@article{salucci2020paradigms,
  title={Paradigms and scenarios for the dark matter phenomenon},
  author={Salucci, Paolo and Turini, Nicola and Di Paolo, Chiara},
  journal={Universe},
  volume={6},
  number={8},
  pages={118},
  year={2020},
  publisher={MDPI}
}

@article{freeman1970disks,
  title={On the disks of spiral and S0 galaxies},
  author={Freeman, Kenneth C},
  journal={Astrophysical Journal, vol. 160, p. 811},
  volume={160},
  pages={811},
  year={1970}
}

@article{salucci2000dark,
  title={Dark matter scaling relations},
  author={Salucci, Paolo and Burkert, Andreas},
  journal={The Astrophysical Journal},
  volume={537},
  number={1},
  pages={L9},
  year={2000},
  publisher={IOP Publishing}
}

@article{nesti2023quest,
  title={The Quest for the Nature of the Dark Matter: The Need of a New Paradigm},
  author={Nesti, Fabrizio and Salucci, Paolo and Turini, Nicola},
  journal={Astronomy},
  volume={2},
  number={2},
  pages={90--104},
  year={2023},
  publisher={MDPI}
}

@article{maccio2020nihao,
  title={NIHAO--XXIII. Dark matter density shaped by black hole feedback},
  author={Macci{\`o}, Andrea V and Crespi, Samuele and Blank, Marvin and Kang, Xi},
  journal={Monthly Notices of the Royal Astronomical Society: Letters},
  volume={495},
  number={1},
  pages={L46--L50},
  year={2020},
  publisher={Oxford University Press}
}

@article{tollet2016nihao,
  title={NIHAO--IV: core creation and destruction in dark matter density profiles across cosmic time},
  author={Tollet, Edouard and Macci{\`o}, Andrea V and Dutton, Aaron A and Stinson, Greg S and Wang, Liang and Penzo, Camilla and Gutcke, Thales A and Buck, Tobias and Kang, Xi and Brook, Chris and others},
  journal={Monthly Notices of the Royal Astronomical Society},
  volume={456},
  number={4},
  pages={3542--3552},
  year={2016},
  publisher={Oxford University Press}
}

@article{karukes2017universal,
  title={The universal rotation curve of dwarf disc galaxies},
  author={Karukes, Ekaterina V and Salucci, Paolo},
  journal={Monthly Notices of the Royal Astronomical Society},
  volume={465},
  number={4},
  pages={4703--4722},
  year={2017},
  publisher={Oxford University Press}
}

@article{donato2004cores,
  title={Cores of dark matter haloes correlate with stellar scalelengths},
  author={Donato, Fiorenza and Gentile, Gianfranco and Salucci, Paolo},
  journal={Monthly Notices of the Royal Astronomical Society},
  volume={353},
  number={2},
  pages={L17--L22},
  year={2004},
  publisher={Blackwell Science Ltd}
}

@article{del2017small,
  title={Small scale problems of the $\Lambda$CDM model: a short review},
  author={Del Popolo, Antonino and Le Delliou, Morgan},
  journal={Galaxies},
  volume={5},
  number={1},
  pages={17},
  year={2017},
  publisher={MDPI}
}

@article{mcgaugh2000baryonic,
  title={The baryonic tully-fisher relation},
  author={McGaugh, Stacy S and Schombert, Jim M and Bothun, Greg D and De Blok, WJG},
  journal={The Astrophysical Journal},
  volume={533},
  number={2},
  pages={L99},
  year={2000},
  publisher={IOP Publishing}
}

@article{girelli2020stellar,
  title={The stellar-to-halo mass relation over the past 12 Gyr-I. Standard $\Lambda$CDM model},
  author={Girelli, Giacomo and Pozzetti, Lucia and Bolzonella, Micol and Giocoli, Carlo and Marulli, Federico and Baldi, Marco},
  journal={Astronomy \& Astrophysics},
  volume={634},
  pages={A135},
  year={2020},
  publisher={EDP Sciences}
}

@article{brook2014stellar,
  title={The stellar-to-halo mass relation for local group galaxies},
  author={Brook, CB and Di Cintio, A and Knebe, A and Gottl{\"o}ber, S and Hoffman, Y and Yepes, G and Garrison-Kimmel, S},
  journal={The Astrophysical Journal Letters},
  volume={784},
  number={1},
  pages={L14},
  year={2014},
  publisher={IOP Publishing}
}

@article{kewley2008metallicity,
  title={Metallicity calibrations and the mass-metallicity relation for star-forming galaxies},
  author={Kewley, Lisa J and Ellison, Sara L},
  journal={The Astrophysical Journal},
  volume={681},
  number={2},
  pages={1183},
  year={2008},
  publisher={IOP Publishing}
}

@article{ma2016origin,
  title={The origin and evolution of the galaxy mass--metallicity relation},
  author={Ma, Xiangcheng and Hopkins, Philip F and Faucher-Gigu{\`e}re, Claude-Andr{\'e} and Zolman, Nick and Muratov, Alexander L and Kere{\v{s}}, Du{\v{s}}an and Quataert, Eliot},
  journal={Monthly Notices of the Royal Astronomical Society},
  volume={456},
  number={2},
  pages={2140--2156},
  year={2016},
  publisher={The Royal Astronomical Society}
}

@article{navarro2000dark,
  title={Dark halo and disk galaxy scaling laws},
  author={Navarro, Julio F},
  journal={arXiv preprint astro-ph/0012334},
  year={2000}
}

@article{dutton2016nihao,
  title={NIHAO V: too big does not fail--reconciling the conflict between $\Lambda$CDM predictions and the circular velocities of nearby field galaxies},
  author={Dutton, Aaron A and Macci{\`o}, Andrea V and Frings, Jonas and Wang, Liang and Stinson, Gregory S and Penzo, Camilla and Kang, Xi},
  journal={Monthly Notices of the Royal Astronomical Society: Letters},
  volume={457},
  number={1},
  pages={L74--L78},
  year={2016},
  publisher={The Royal Astronomical Society}
}

@article{buck2020nihao,
  title={NIHAO-UHD: the properties of MW-like stellar discs in high-resolution cosmological simulations},
  author={Buck, Tobias and Obreja, Aura and Macci{\`o}, Andrea V and Minchev, Ivan and Dutton, Aaron A and Ostriker, Jeremiah P},
  journal={Monthly Notices of the Royal Astronomical Society},
  volume={491},
  number={3},
  pages={3461--3478},
  year={2020},
  publisher={Oxford University Press}
}

@article{tully1977new,
  title={A new method of determining distances to galaxies},
  author={Tully, R Brent and Fisher, J Richard},
  journal={Astronomy and Astrophysics, vol. 54, no. 3, Feb. 1977, p. 661-673.},
  volume={54},
  pages={661--673},
  year={1977}
}

@article{lelli2013scaling,
  title={A scaling relation for disc galaxies: Circular-velocity gradient versus central surface brightness},
  author={Lelli, Federico and Fraternali, Filippo and Verheijen, Marc},
  journal={Monthly Notices of the Royal Astronomical Society: Letters},
  volume={433},
  number={1},
  pages={L30--L34},
  year={2013},
  publisher={Oxford University Press}
}

@article{pizzella2005relation,
  title={On the relation between circular velocity and central velocity dispersion in high and low surface brightness galaxies},
  author={Pizzella, Alessandro and Corsini, EM and Dalla Bont{\`a}, Elena and Sarzi, M and Coccato, Lodovico and Bertola, Francesco},
  journal={The Astrophysical Journal},
  volume={631},
  number={2},
  pages={785},
  year={2005},
  publisher={IOP Publishing}
}

@article{di2019universal,
  title={The universal rotation curve of low surface brightness galaxies--IV. The interrelation between dark and luminous matter},
  author={Di Paolo, Chiara and Salucci, Paolo and Erkurt, Adnan},
  journal={Monthly Notices of the Royal Astronomical Society},
  volume={490},
  number={4},
  pages={5451--5477},
  year={2019},
  publisher={Oxford University Press}
}

@article{lelli2017one,
  title={One law to rule them all: the radial acceleration relation of galaxies},
  author={Lelli, Federico and McGaugh, Stacy S and Schombert, James M and Pawlowski, Marcel S},
  journal={The Astrophysical Journal},
  volume={836},
  number={2},
  pages={152},
  year={2017},
  publisher={IOP Publishing}
}

@article{de2022accurate,
  title={The accurate mass distribution of M87, the Giant Galaxy with imaged shadow of its supermassive black hole, as a portal to new Physics},
  author={De Laurentis, Mariafelicia and Salucci, Paolo},
  journal={The Astrophysical Journal},
  volume={929},
  number={1},
  pages={17},
  year={2022},
  publisher={IOP Publishing}
}

@article{keller2019chaos,
  title={Chaos and variance in galaxy formation},
  author={Keller, BW and Wadsley, JW and Wang, L and Kruijssen, JM Diederik},
  journal={Monthly Notices of the Royal Astronomical Society},
  volume={482},
  number={2},
  pages={2244--2261},
  year={2019},
  publisher={Oxford University Press}
}

@article{cole1994recipe,
  title={A recipe for galaxy formation},
  author={Cole, Shaun and Aragon-Salamanca, Alfonso and Frenk, Carlos S and Navarro, Julio F and Zepf, Stephen E},
  journal={Monthly Notices of the Royal Astronomical Society},
  volume={271},
  number={4},
  pages={781--806},
  year={1994},
  publisher={Oxford University Press Oxford, UK}
}

@article{maccio2007concentration,
  title={Concentration, spin and shape of dark matter haloes: scatter and the dependence on mass and environment},
  author={Macci\`o, Andrea V and Dutton, Aaron A and Van Den Bosch, Frank C and Moore, Ben and Potter, Doug and Stadel, Joachim},
  journal={Monthly Notices of the Royal Astronomical Society},
  volume={378},
  number={1},
  pages={55--71},
  year={2007},
  publisher={Blackwell Publishing Ltd Oxford, UK}
}

@article{foot2014tully,
  title={Tully-Fisher relation, galactic rotation curves and dissipative mirror dark matter},
  author={Foot, R},
  journal={Journal of Cosmology and Astroparticle Physics},
  volume={2014},
  number={12},
  pages={047},
  year={2014},
  publisher={IOP Publishing}
}

@ARTICLE{buck2021,
       author = {{Buck}, Tobias and {Rybizki}, Jan and {Buder}, Sven and {Obreja}, Aura and {Macci{\`o}}, Andrea V. and {Pfrommer}, Christoph and {Steinmetz}, Matthias and {Ness}, Melissa},
        title = "{The challenge of simultaneously matching the observed diversity of chemical abundance patterns in cosmological hydrodynamical simulations}",
      journal = {\mnras},
         year = 2021,
        month = dec,
       volume = {508},
       number = {3},
        pages = {3365-3387},
          doi = {10.1093/mnras/stab2736},
archivePrefix = {arXiv},
       eprint = {2103.03884},
 primaryClass = {astro-ph.GA},
       adsurl = {https://ui.adsabs.harvard.edu/abs/2021MNRAS.508.3365B}
}

@article{maccio2016nihao,
  title={NIHAO X: reconciling the local galaxy velocity function with cold dark matter via mock H i observations},
  author={Macci{\`o}, Andrea V and Udrescu, Silviu M and Dutton, Aaron A and Obreja, Aura and Wang, Liang and Stinson, Greg R and Kang, Xi},
  journal={Monthly Notices of the Royal Astronomical Society: Letters},
  volume={463},
  number={1},
  pages={L69--L73},
  year={2016},
  publisher={The Royal Astronomical Society}
}

@article{DeRossi2017,
  author  = {Mar{\'\i}a Emilia De Rossi and Richard G. Bower and Andreea S. Font and Joop Schaye and Tom Theuns},
  title   = {Galaxy metallicity scaling relations in the EAGLE simulations},
  journal = {Monthly Notices of the Royal Astronomical Society},
  year    = {2017},
  volume  = {472},
  number  = {3},
  pages   = {3354--3367},
  doi     = {10.1093/mnras/stx2240}
}

@ARTICLE{maccio2020,
       author = {{Macci{\`o}}, Andrea V. and {Courteau}, St{\'e}phane and {Ouellette}, Nathalie N.-Q. and {Dutton}, Aaron A.},
        title = "{Abundance matching tested on small scales with galaxy dynamics}",
      journal = {\mnras},
         year = 2020,
        month = jul,
       volume = {496},
       number = {1},
        pages = {L101-L105},
          doi = {10.1093/mnrasl/slaa094},
archivePrefix = {arXiv},
       eprint = {2006.00818},
 primaryClass = {astro-ph.GA},
       adsurl = {https://ui.adsabs.harvard.edu/abs/2020MNRAS.496L.101M}
}

@article{Nelson2019,
  author  = {Dylan Nelson and Volker Springel and Federico Marinacci and Paul Torrey and Mark Vogelsberger and Annalisa Pillepich and …},
  title   = {The IllustrisTNG simulations: public data release},
  journal = {Computational Astrophysics and Cosmology},
  year    = {2019},
  volume  = {6},
  number  = {2},
  doi     = {10.1186/s40668-019-0028-x}
}

@ARTICLE{abadi2010,
       author = {{Abadi}, Mario G. and {Navarro}, Julio F. and {Fardal}, Mark and {Babul}, Arif and {Steinmetz}, Matthias},
        title = "{Galaxy-induced transformation of dark matter haloes}",
      journal = {\mnras},
         year = 2010,
        month = sep,
       volume = {407},
       number = {1},
        pages = {435-446},
          doi = {10.1111/j.1365-2966.2010.16912.x},
archivePrefix = {arXiv},
       eprint = {0902.2477},
 primaryClass = {astro-ph.GA},
       adsurl = {https://ui.adsabs.harvard.edu/abs/2010MNRAS.407..435A}
}

@ARTICLE{Speagle2020MNRAS,
       author = {{Speagle}, Joshua S.},
        title = "{DYNESTY: a dynamic nested sampling package for estimating Bayesian posteriors and evidences}",
      journal = {\mnras},
         year = 2020,
        month = apr,
       volume = {493},
       number = {3},
        pages = {3132-3158},
          doi = {10.1093/mnras/staa278},
archivePrefix = {arXiv},
       eprint = {1904.02180},
 primaryClass = {astro-ph.IM},
       adsurl = {https://ui.adsabs.harvard.edu/abs/2020MNRAS.493.3132S}
}

@MISC{Koposov2023zndo,
       author = {{Koposov}, Sergey and {Speagle}, Josh and {Barbary}, Kyle and {Ashton}, Gregory and {Bennett}, Ed and {Buchner}, Johannes and {Scheffler}, Carl and {Cook}, Ben and {Talbot}, Colm and {Guillochon}, James and {Cubillos}, Patricio and {Asensio Ramos}, Andr{\'e}s and {Johnson}, Ben and {Lang}, Dustin and {Ilya} and {Dartiailh}, Matthieu and {Nitz}, Alex and {McCluskey}, Andrew and {Archibald}, Anne},
        title = "{joshspeagle/dynesty: v2.1.3}",
         year = 2023,
        month = oct,
          eid = {10.5281/zenodo.8408702},
          doi = {10.5281/zenodo.8408702},
      version = {v2.1.3},
    publisher = {Zenodo},
       adsurl = {https://ui.adsabs.harvard.edu/abs/2023zndo...8408702K}
}

@ARTICLE{Blumenthal1986AB,
       author = {{Blumenthal}, G.~R. and {Faber}, S.~M. and {Flores}, R. and {Primack}, J.~R.},
        title = "{Contraction of Dark Matter Galactic Halos Due to Baryonic Infall}",
      journal = {\apj},
         year = 1986,
        month = feb,
       volume = {301},
        pages = {27},
          doi = {10.1086/163867},
       adsurl = {https://ui.adsabs.harvard.edu/abs/1986ApJ...301...27B}
}

@ARTICLE{maccio2012ApJ,
       author = {{Macci{\`o}}, A.~V. and {Stinson}, G. and {Brook}, C.~B. and {Wadsley}, J. and {Couchman}, H.~M.~P. and {Shen}, S. and {Gibson}, B.~K. and {Quinn}, T.},
        title = "{Halo Expansion in Cosmological Hydro Simulations: Toward a Baryonic Solution of the Cusp/Core Problem in Massive Spirals}",
      journal = {\apjl},
         year = 2012,
        month = jan,
       volume = {744},
       number = {1},
          eid = {L9},
        pages = {L9},
          doi = {10.1088/2041-8205/744/1/L9},
archivePrefix = {arXiv},
       eprint = {1111.5620},
 primaryClass = {astro-ph.CO},
       adsurl = {https://ui.adsabs.harvard.edu/abs/2012ApJ...744L...9M}
}

@ARTICLE{Brook2011,
       author = {{Brook}, C.~B. and {Governato}, F. and {Ro{\v{s}}kar}, R. and {Stinson}, G. and {Brooks}, A.~M. and {Wadsley}, J. and {Quinn}, T. and {Gibson}, B.~K. and {Snaith}, O. and {Pilkington}, K. and {House}, E. and {Pontzen}, A.},
        title = "{Hierarchical formation of bulgeless galaxies: why outflows have low angular momentum}",
      journal = {\mnras},
         year = 2011,
        month = aug,
       volume = {415},
       number = {2},
        pages = {1051-1060},
          doi = {10.1111/j.1365-2966.2011.18545.x},
archivePrefix = {arXiv},
       eprint = {1010.1004},
 primaryClass = {astro-ph.CO},
       adsurl = {https://ui.adsabs.harvard.edu/abs/2011MNRAS.415.1051B}
}

@ARTICLE{Pontzen2012,
       author = {{Pontzen}, Andrew and {Governato}, Fabio},
        title = "{How supernova feedback turns dark matter cusps into cores}",
      journal = {\mnras},
         year = 2012,
        month = apr,
       volume = {421},
       number = {4},
        pages = {3464-3471},
          doi = {10.1111/j.1365-2966.2012.20571.x},
archivePrefix = {arXiv},
       eprint = {1106.0499},
 primaryClass = {astro-ph.CO},
       adsurl = {https://ui.adsabs.harvard.edu/abs/2012MNRAS.421.3464P}
}

@ARTICLE{chan2015,
       author = {{Chan}, T.~K. and {Kere{\v{s}}}, D. and {O{\~n}orbe}, J. and {Hopkins}, P.~F. and {Muratov}, A.~L. and {Faucher-Gigu{\`e}re}, C.-A. and {Quataert}, E.},
        title = "{The impact of baryonic physics on the structure of dark matter haloes: the view from the FIRE cosmological simulations}",
      journal = {\mnras},
         year = 2015,
        month = dec,
       volume = {454},
       number = {3},
        pages = {2981-3001},
          doi = {10.1093/mnras/stv2165},
archivePrefix = {arXiv},
       eprint = {1507.02282},
 primaryClass = {astro-ph.GA},
       adsurl = {https://ui.adsabs.harvard.edu/abs/2015MNRAS.454.2981C}
}

@article{Vogelsberger2020,
  author  = {Mark Vogelsberger and Federico Marinacci and Paul Torrey and et al.},
  title   = {Cosmological simulations of galaxy formation},
  journal = {Nature Reviews Physics},
  year    = {2020},
  volume  = {2},
  pages   = {42–66},
  doi     = {10.1038/s42254-019-0127-2}
}

@article{Crain2015,
  author  = {Robert A. Crain and Joop Schaye and Richard G. Bower and Michelle Furlong and Matthieu Schaller and Tom Theuns and Claudio Dalla Vecchia and Carlos S. Frenk and Ian G. McCarthy and John C. Helly and Adrian Jenkins and Yetli M. Rosas-Guevara and Simon D. M. White and James W. Trayford},
  title   = {The EAGLE simulations of galaxy formation: calibration of subgrid physics and model variations},
  journal = {Monthly Notices of the Royal Astronomical Society},
  year    = {2015},
  volume  = {450},
  number  = {2},
  pages   = {1937–1961},
  doi     = {10.1093/mnras/stv832}
}

@article{dutton2017nihao,
  title={NIHAO XII: galactic uniformity in a $\Lambda$CDM universe},
  author={Dutton, Aaron A and Obreja, Aura and Wang, Liang and Gutcke, Thales A and Buck, Tobias and Udrescu, Silviu M and Frings, Jonas and Stinson, Gregory S and Kang, Xi and Macci{\`o}, Andrea V},
  journal={Monthly Notices of the Royal Astronomical Society},
  volume={467},
  number={4},
  pages={4937--4950},
  year={2017},
  publisher={Oxford University Press}
}

@article{santos2018nihao,
  title={NIHAO--XIV. Reproducing the observed diversity of dwarf galaxy rotation curve shapes in $\Lambda$CDM},
  author={Santos-Santos, Isabel M and Di Cintio, Arianna and Brook, Chris B and Macci{\`o}, Andrea and Dutton, Aaron and Dom{\'\i}nguez-Tenreiro, Rosa},
  journal={Monthly Notices of the Royal Astronomical Society},
  volume={473},
  number={4},
  pages={4392--4403},
  year={2018},
  publisher={Oxford University Press}
}

@article{buck2019nihao,
  title={NIHAO XV: the environmental impact of the host galaxy on galactic satellite and field dwarf galaxies},
  author={Buck, Tobias and Macci{\`o}, Andrea V and Dutton, Aaron A and Obreja, Aura and Frings, Jonas},
  journal={Monthly Notices of the Royal Astronomical Society},
  volume={483},
  number={1},
  pages={1314--1341},
  year={2019},
  publisher={Oxford University Press}
}

@ARTICLE{Cannarozzo2020MNRAS,
       author = {{Cannarozzo}, Carlo and {Sonnenfeld}, Alessandro and {Nipoti}, Carlo},
        title = "{The cosmic evolution of the stellar mass-velocity dispersion relation of early-type galaxies}",
      journal = {\mnras},
         year = 2020,
        month = oct,
       volume = {498},
       number = {1},
        pages = {1101-1120},
          doi = {10.1093/mnras/staa2147},
archivePrefix = {arXiv},
       eprint = {1910.06987},
 primaryClass = {astro-ph.GA},
       adsurl = {https://ui.adsabs.harvard.edu/abs/2020MNRAS.498.1101C}
}

@ARTICLE{Waterval2025MNRAS,
       author = {{Waterval}, Stefan and {Cannarozzo}, Carlo and {Macci{\`o}}, Andrea V.},
        title = "{Gas accretion at high redshift: cold flows all the way}",
      journal = {\mnras},
         year = 2025,
        month = mar,
       volume = {537},
       number = {3},
        pages = {2726-2751},
          doi = {10.1093/mnras/staf198},
archivePrefix = {arXiv},
       eprint = {2501.19009},
 primaryClass = {astro-ph.GA},
       adsurl = {https://ui.adsabs.harvard.edu/abs/2025MNRAS.537.2726W}
}

@article{arora2023manga,
  title={MaNGA galaxy properties--II. A detailed comparison of observed and simulated spiral galaxy scaling relations},
  author={Arora, Nikhil and Courteau, St{\'e}phane and Stone, Connor and Macci{\`o}, Andrea V},
  journal={Monthly Notices of the Royal Astronomical Society},
  volume={522},
  number={1},
  pages={1208--1227},
  year={2023},
  publisher={Oxford University Press}
}

@article{begeman1991extended,
  title={Extended rotation curves of spiral galaxies: Dark haloes and modified dynamics},
  author={Begeman, KG and Broeils, AH and Sanders, RH},
  journal={Monthly Notices of the Royal Astronomical Society},
  volume={249},
  number={3},
  pages={523--537},
  year={1991},
  publisher={The Royal Astronomical Society}
}

@article{burkert1995structure,
  title={The structure of dark matter halos in dwarf galaxies},
  author={Burkert, Andreas},
  journal={The Astrophysical Journal},
  volume={447},
  number={1},
  pages={L25},
  year={1995},
  publisher={IOP publishing}
}

@article{somerville2015physical,
  title={Physical models of galaxy formation in a cosmological framework},
  author={Somerville, Rachel S and Dav{\'e}, Romeel},
  journal={Annual Review of Astronomy and Astrophysics},
  volume={53},
  number={1},
  pages={51--113},
  year={2015},
  publisher={Annual Reviews}
}

@article{naab2017theoretical,
  title={Theoretical challenges in galaxy formation},
  author={Naab, Thorsten and Ostriker, Jeremiah P},
  journal={Annual review of astronomy and astrophysics},
  volume={55},
  number={1},
  pages={59--109},
  year={2017},
  publisher={Annual Reviews}
}

@article{hopkins2018fire,
  title={FIRE-2 simulations: physics versus numerics in galaxy formation},
  author={Hopkins, Philip F and Wetzel, Andrew and Kere{\v{s}}, Du{\v{s}}an and Faucher-Gigu{\`e}re, Claude-Andr{\'e} and Quataert, Eliot and Boylan-Kolchin, Michael and Murray, Norman and Hayward, Christopher C and Garrison-Kimmel, Shea and Hummels, Cameron and others},
  journal={Monthly Notices of the Royal Astronomical Society},
  volume={480},
  number={1},
  pages={800--863},
  year={2018},
  publisher={Oxford University Press}
}

@article{conselice2014evolution,
  title={The evolution of galaxy structure over cosmic time},
  author={Conselice, Christopher J},
  journal={Annual Review of Astronomy and Astrophysics},
  volume={52},
  number={1},
  pages={291--337},
  year={2014},
  publisher={Annual Reviews}
}

@article{torrey2015synthetic,
  title={Synthetic galaxy images and spectra from the Illustris simulation},
  author={Torrey, Paul and Snyder, Gregory F and Vogelsberger, Mark and Hayward, Christopher C and Genel, Shy and Sijacki, Debora and Springel, Volker and Hernquist, Lars and Nelson, Dylan and Kriek, Mariska and others},
  journal={Monthly Notices of the Royal Astronomical Society},
  volume={447},
  number={3},
  pages={2753--2771},
  year={2015},
  publisher={Oxford University Press}
}

@article{schaye2015eagle,
  title={The EAGLE project: simulating the evolution and assembly of galaxies and their environments},
  author={Schaye, Joop and Crain, Robert A and Bower, Richard G and Furlong, Michelle and Schaller, Matthieu and Theuns, Tom and Dalla Vecchia, Claudio and Frenk, Carlos S and McCarthy, IG and Helly, John C and others},
  journal={Monthly Notices of the Royal Astronomical Society},
  volume={446},
  number={1},
  pages={521--554},
  year={2015},
  publisher={Oxford University Press}
}

@article{pillepich2018first,
  title={First results from the IllustrisTNG simulations: the stellar mass content of groups and clusters of galaxies},
  author={Pillepich, Annalisa and Nelson, Dylan and Hernquist, Lars and Springel, Volker and Pakmor, R{\"u}diger and Torrey, Paul and Weinberger, Rainer and Genel, Shy and Naiman, Jill P and Marinacci, Federico and others},
  journal={Monthly Notices of the Royal Astronomical Society},
  volume={475},
  number={1},
  pages={648--675},
  year={2018},
  publisher={Oxford University Press}
}

@article{rodriguez2019optical,
  title={The optical morphologies of galaxies in the IllustrisTNG simulation: a comparison to Pan-STARRS observations},
  author={Rodriguez-Gomez, Vicente and Snyder, Gregory F and Lotz, Jennifer M and Nelson, Dylan and Pillepich, Annalisa and Springel, Volker and Genel, Shy and Weinberger, Rainer and Tacchella, Sandro and Pakmor, R{\"u}diger and others},
  journal={Monthly Notices of the Royal Astronomical Society},
  volume={483},
  number={3},
  pages={4140--4159},
  year={2019},
  publisher={Oxford University Press}
}

@article{conroy2009propagation,
  title={The propagation of uncertainties in stellar population synthesis modeling. I. The relevance of uncertain aspects of stellar evolution and the initial mass function to the derived physical properties of galaxies},
  author={Conroy, Charlie and Gunn, James E and White, Martin},
  journal={The Astrophysical Journal},
  volume={699},
  number={1},
  pages={486},
  year={2009},
  publisher={IOP Publishing}
}

@ARTICLE{Bullock01,
       author = {{Bullock}, J.~S. and {Kolatt}, T.~S. and {Sigad}, Y. and {Somerville}, R.~S. and {Kravtsov}, A.~V. and {Klypin}, A.~A. and {Primack}, J.~R. and {Dekel}, A.},
        title = "{Profiles of dark haloes: evolution, scatter and environment}",
      journal = {\mnras},
         year = 2001,
        month = mar,
       volume = {321},
       number = {3},
        pages = {559-575},
          doi = {10.1046/j.1365-8711.2001.04068.x},
archivePrefix = {arXiv},
       eprint = {astro-ph/9908159},
 primaryClass = {astro-ph},
       adsurl = {https://ui.adsabs.harvard.edu/abs/2001MNRAS.321..559B}
}

@ARTICLE{MoMaoWhite1998,
  author       = {Mo, H. J. and Mao, Shude and White, Simon D. M.},
  title        = {The formation of galactic discs},
  journal      = {Monthly Notices of the Royal Astronomical Society},
  year         = {1998},
  volume       = {295},
  number       = {2},
  pages        = {319--336},
  doi          = {10.1046/j.1365-8711.1998.01227.x}
}

@article{courteau2007bulge,
  title={The bulge-halo connection in galaxies: A physical interpretation of the vc-$\sigma$0 relation},
  author={Courteau, St{\'e}phane and McDonald, Michael and Widrow, Lawrence M and Holtzman, Jon},
  journal={The Astrophysical Journal},
  volume={655},
  number={1},
  pages={L21},
  year={2007},
  publisher={IOP Publishing}
}

@article{stone2019intrinsic,
  title={The intrinsic scatter of the radial acceleration relation},
  author={Stone, Connor and Courteau, St{\'e}phane},
  journal={The Astrophysical Journal},
  volume={882},
  number={1},
  pages={6},
  year={2019},
  publisher={IOP Publishing}
}

@article{dutton2007revised,
  title={A revised model for the formation of disk galaxies: low spin and dark halo expansion},
  author={Dutton, Aaron A and van den Bosch, Frank C and Dekel, Avishai and Courteau, St{\'e}phane},
  journal={The Astrophysical Journal},
  volume={654},
  number={1},
  pages={27},
  year={2007},
  publisher={IOP Publishing}
}

@article{frosst2022diversity,
  title={The diversity of spiral galaxies explained},
  author={Frosst, Matthew and Courteau, St{\'e}phane and Arora, Nikhil and Stone, Connor and Macci{\`o}, Andrea V and Blank, Marvin},
  journal={Monthly Notices of the Royal Astronomical Society},
  volume={514},
  number={3},
  pages={3510--3531},
  year={2022},
  publisher={Oxford University Press}
}

@article{macarthur2003structure,
  title={Structure of disk-dominated galaxies. I. Bulge/disk parameters, simulations, and secular evolution},
  author={MacArthur, Lauren A and Courteau, Stephane and Holtzman, Jon A},
  journal={The Astrophysical Journal},
  volume={582},
  number={2},
  pages={689},
  year={2003},
  publisher={IOP Publishing}
}



\appendix

\section{Burkert versus ISO fits} \label{appendix}
We fit the dark matter density profiles of the galaxies using both the isothermal (ISO) and Burkert models, and computed the corresponding $\chi^2$ values for each fit. On average, the Burkert profile yields $\chi^2$ values approximately 48\% higher than those of the ISO model (Fig. \ref{chi2}), indicating a systematically poorer fit. The fitting procedure was restricted to the central dark matter region, typically spanning 10–50 kpc, depending on the size of each galaxy. Notably, the Burkert profile begins to deviate significantly from the simulated density distribution beyond roughly three-quarters of this radial range. This divergence arises because the Burkert profile declines more steeply with radius ($\propto r^{-3}$) compared to the isothermal profile ($\propto r^{-2}$).

\begin{figure}
    \includegraphics[width=\columnwidth]{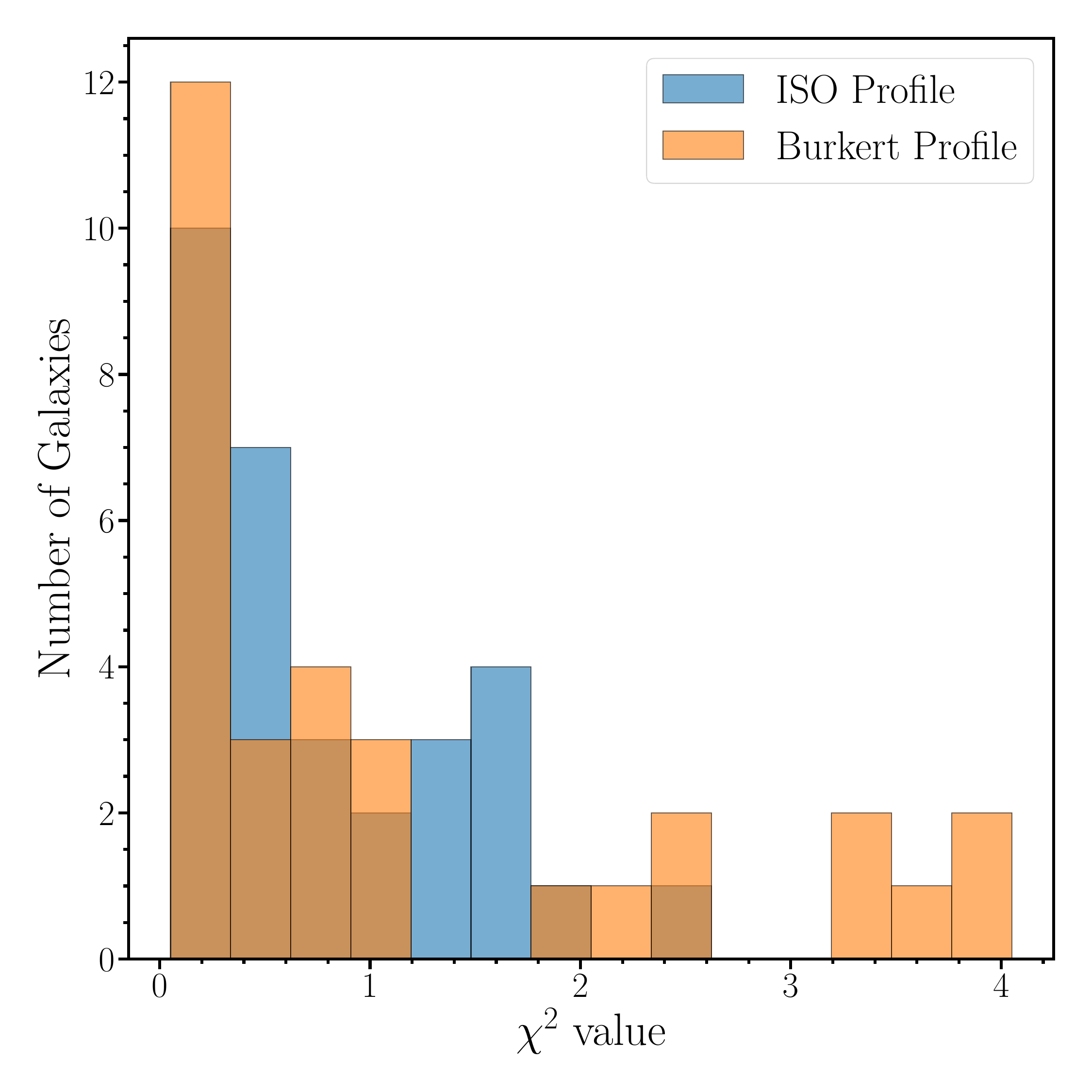}
    \caption{Histograms of the $\chi^2$ values for dark matter fits using the Burkert profile versus the fits using the ISO profile. The Burkert profile shows higher values on average.}
    \label{chi2}
\end{figure}

\section{The Bayesian hierarchical framework}
\label{app:model}

To infer the redshift evolution of the $R_d{-}r_0$ relation and its intrinsic scatter, we employ a Bayesian hierarchical framework. This approach allows us to describe the global behaviour of the simulated galaxy population through a set of hyperparameters that govern the evolution of the relation across cosmic time.

Each galaxy in the NIHAO sample is characterized by three physical quantities:
\begin{itemize}
    \item the disc scale radius, $R_d$;
    \item the DM scale radius, $r_0$;
    \item the redshift, $z$.
\end{itemize}

We collect these quantities into a vector of parameters
$\boldsymbol{\Phi} = \{R_d,\, r_0,\, z\}$.
Under the hierarchical assumption, the galaxies are considered realizations drawn from a common parent distribution described by a set of hyperparameters $\boldsymbol{\Theta}$:
\begin{equation}
\mathcal{P}(\boldsymbol{\Phi}) = \mathcal{P}(\boldsymbol{\Phi}\,|\,\boldsymbol{\Theta}).
\end{equation}

According to Bayes’ theorem, the posterior probability of the hyperparameters given the data, $\mathcal{D}$, is
\begin{equation}
\mathcal{P}(\boldsymbol{\Theta}\,|\,\mathcal{D}) =
\frac{\mathcal{P}(\mathcal{D}\,|\,\boldsymbol{\Theta})\,
      \mathcal{P}(\boldsymbol{\Theta})}
     {\mathcal{P}(\mathcal{D})},
\label{eq:bayes}
\end{equation}
where $\mathcal{P}(\mathcal{D}\,|\,\boldsymbol{\Theta})$ is the \emph{likelihood}, 
$\mathcal{P}(\boldsymbol{\Theta})$ represents the \emph{prior} distributions of the hyperparameters, and $\mathcal{P}(\mathcal{D})$ is the \emph{Bayesian evidence} ensuring proper normalization.

Since the values of $\{R_d^\mathrm{sim},\, r_0^\mathrm{sim}\, z^\mathrm{sim}\}$ are directly obtained from the simulation outputs and thus are free from observational uncertainties, the likelihood for the $i$-th galaxy simplifies to
\begin{equation}
\begin{aligned}
\mathcal{P}(\mathcal{D}_i\,|\,\boldsymbol{\Theta})
&= \int \mathcal{P}(\mathcal{D}_i\,|\,\boldsymbol{\Phi}_i)\,
         \mathcal{P}(\boldsymbol{\Phi}_i\,|\,\boldsymbol{\Theta})\,
         \mathrm{d}\boldsymbol{\Phi}_i =\\
&= \int \delta(\mathcal{D}_i - \boldsymbol{\Phi}_i)\,
         \mathcal{P}(\boldsymbol{\Phi}_i\,|\,\boldsymbol{\Theta})\,
         \mathrm{d}\boldsymbol{\Phi}_i.
\end{aligned}
\label{eq:likelihood}
\end{equation}

The probability distribution describing the $R_d{-}r_0$ relation at a given redshift is modeled as a normal distribution,
\begin{equation}
\mathcal{P}(\boldsymbol{\Phi}\,|\,\boldsymbol{\Theta})
= \mathcal{N}\, \Big(\log r_0 \,\big|\, 
      \mu(z, \log R_d),\, \sigma^2(z)\Big),
\label{eq:normal}
\end{equation}
where $\mu(z, \log R_d)$ is the mean relation defined in \autoref{eq:mean}, 
and $\sigma(z)$ represents the intrinsic scatter of the population as in \autoref{eq:scatter}.  

The posterior distributions of all hyperparameters
\[
\boldsymbol{\Theta} = 
\{\log r_{0,0},\, \log r_{0,z},\, \alpha_0,\, \alpha_z,\, \sigma_0,\, \sigma_z\},
\]
are sampled using the nested sampling algorithm implemented in \textsc{dynesty} \citep{Speagle2020MNRAS,Koposov2023zndo}, adopting the configuration summarized in \autoref{tab:dynesty_settings}. 
The sampler was initialized with a fixed random state to ensure reproducibility and run in 
dynamic mode, which adaptively allocates new live points during the exploration of the posterior volume.
This technique efficiently explores multi-dimensional parameter spaces and provides both the full posterior and the Bayesian evidence.
\begin{table}
\centering
\caption{Settings adopted for the \textsc{dynesty} nested sampling run. 
Column 1 lists the parameters of the \texttt{DynamicNestedSampler} class and the 
\texttt{run\_nested} method. Column 2 reports the adopted values.}
\begin{tabular}{lc}
\hline
Parameter & Value \\
\hline
\texttt{nlive}        & 1000 \\
\texttt{dlogz\_init}  & 0.01 \\
\texttt{sampler}      & \texttt{DynamicNestedSampler} \\
\texttt{rstate}       & \texttt{numpy.random.default\_rng(18)} \\
\texttt{ndim}         & \texttt{len(param\_labels)} \\
\hline
\end{tabular}
\label{tab:dynesty_settings}
\end{table}
\autoref{tab:priors} lists the priors adopted for the model.
\begin{table}
\centering
\caption{Prior ranges adopted for the hyperparameters of the evolutionary single power-law model.}
\begin{tabular}{lcc}
\hline
Parameter & Description & Prior range \\
\hline
$\log r_{0,0}$ & Normalization at $z=0$ & $\mathcal{U}(0.0,\;1.5)$ \\
$\log r_{0,z}$ & Redshift dependence of normalization & $\mathcal{U}(-0.5,\;0.5)$ \\
$\alpha_0$     & Slope at $z=0$ & $\mathcal{U}(0.5,\;3.0)$ \\
$\alpha_z$     & Redshift dependence of slope & $\mathcal{U}(-0.5,\;0.5)$ \\
$\sigma_0$     & Intrinsic scatter at $z=0$ & $\mathcal{U}(0.01,\;0.50)$ \\
$\sigma_z$     & Redshift dependence of scatter & $\mathcal{U}(-0.2,\;0.2)$ \\
\hline
\end{tabular}
\label{tab:priors}
\end{table}
The posterior distributions of all model parameters are shown in \autoref{fig:corner}. 
The contours highlight the covariance structure among the hyperparameters, with diagonal panels displaying the marginalized one-dimensional posterior probability density functions (PDFs).
This hierarchical formalism allows us to self-consistently quantify the redshift evolution of both the slope and normalization of the $R_d{-}r_0$ relation, as well as the gradual tightening of its intrinsic scatter toward lower redshifts, consistent with the trends discussed in \autoref{time_evolution}.

\begin{figure*}
    \includegraphics[width=1.75\columnwidth]{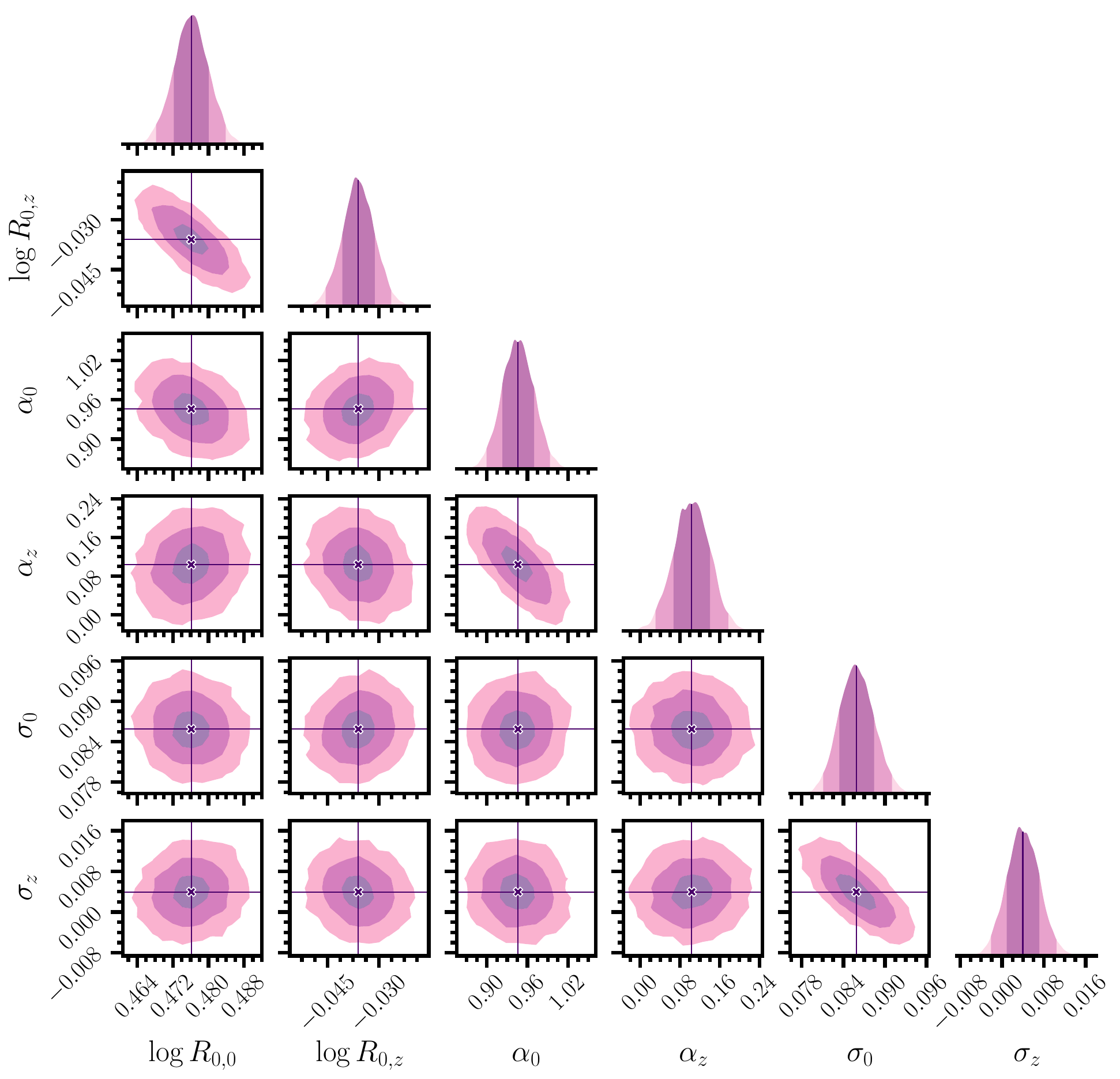}
    \caption{Posterior distributions of the model parameters describing the redshift evolution of the $R_d{-}r_0$ relation.
Diagonal panels show the marginalized one-dimensional probability density functions (PDFs), while off-diagonal panels display the two-dimensional joint posteriors with the 1, 2, and 3$\sigma$ credible contours.
The parameters include the normalization $\log R_{0,0}$ and its redshift evolution term $\log R_{0,z}$ of \autoref{eq:normalization}, the slope $\alpha_0$ and its evolution $\alpha_z$ of \autoref{eq:slope}, and the intrinsic scatter $\sigma_0$ and its evolution $\sigma_z$ of \autoref{eq:scatter}.
Contours and shaded regions correspond to the 68\%, 95\%, and 99.7\% confidence intervals derived from the nested sampling inference.}
    \label{fig:corner}
\end{figure*}


\bsp	
\label{lastpage}
\end{document}